\documentclass[journal, onecolumn, 11pt]{IEEEtran}
\usepackage{hyperref}

\usepackage{ifpdf}

\usepackage{cite}

\ifCLASSINFOpdf
\usepackage[pdftex]{graphicx}
\else
\fi
\usepackage[cmex10]{amsmath}
\usepackage{algorithmic}

\usepackage{array}

\ifCLASSOPTIONcompsoc
\usepackage[caption=false,font=normalsize,labelfont=sf,textfont=sf]{subfig}
\else
\usepackage[caption=false,font=footnotesize]{subfig}
\fi
\usepackage{dblfloatfix}

\ifCLASSOPTIONcaptionsoff
\usepackage[nomarkers]{endfloat}
\let\MYoriglatexcaption\caption
\renewcommand{\caption}[2][\relax]{\MYoriglatexcaption[#2]{#2}}
\fi
\usepackage{url}

\usepackage{amssymb}
\usepackage{color}
\usepackage{gensymb}
\usepackage{bm}
\usepackage{enumerate}

\newcommand{\defi} { \stackrel{\bigtriangleup}{=} }
\newcommand{\mb}{\mathbf}

\def\qed{\hfill\vrule height 1.6ex width 1.5ex depth -.1ex}
\newcommand{\be}{\begin{equation}}
\newcommand{\ee}{\end{equation}}
\newcommand{\ben}{\begin{equation*}}
\newcommand{\een}{\end{equation*}}
\newcommand{\ba}{\begin{array}}
	\newcommand{\ea}{\end{array}}
\newcommand{\rr}{\mathop{{\rm I}\mskip-4.0mu{\rm R}}\nolimits}
\newcommand{\bs}{\boldsymbol}

\newcommand{\Z}{{\mathbb Z}}
\newtheorem{theorem}{Theorem}
\newtheorem{remark}{Remark}

\begin{document}
	%
	\title{Decentralized consensus finite-element \\ Kalman filter for field estimation}
	%
	%
	%
	
	\author{Giorgio~Battistelli,
		Luigi~Chisci,~\IEEEmembership{Senior~Member,~IEEE},
		Nicola~Forti,
		Stefano~Selleri,~\IEEEmembership{Senior~Member,~IEEE},
		and~Giuseppe~Pelosi
		~\IEEEmembership{Fellow,~IEEE}
		\thanks{The authors are with the Dipartimento
			di Ingegneria dell'Informazione (DINFO), Universit\`a di Firenze, Firenze,
			Italy. e-mails: \{giorgio.battistelli,luigi.chisci,nicola.forti,stefano.selleri,giuseppe.pelosi\} @unifi.it.}
}
	
	%
	%

	\markboth{}%
	{Battistelli  \MakeLowercase{\textit{et al.}}: Bare Demo of IEEEtran.cls for Journals}
	%



	\maketitle
	
	\begin{abstract}
		The paper deals with decentralized state estimation for spatially distributed systems described by linear partial differential equations from discrete in-space-and-time noisy measurements provided by sensors deployed over the spatial domain of interest. 
		A fully scalable approach is pursued by decomposing the domain into overlapping subdomains assigned to
		different processing nodes interconnected to form a network.
		Each node runs a local 
		finite-dimensional Kalman filter 
		which exploits the finite element approach for spatial discretization and the parallel Schwarz method to iteratively enforce consensus on the estimates and covariances over the boundaries of adjacent subdomains.
		Stability of the proposed distributed consensus-based finite element Kalman filter is mathematically proved and its effectiveness is demonstrated via simulation experiments concerning the estimation of a bi-dimensional
		temperature field.
	\end{abstract}
	

	%
	\IEEEpeerreviewmaketitle

	\section{Introduction}
	%
	%
	%
	%
	\IEEEPARstart{T}{he} recent breakthrough of wireless sensor network technology has made possible to cost-effectively monitor spatially distributed systems via deployment of
	multiple sensors over the area of interest. 
	This clearly paves the way for several important practical monitoring applications concerning, e.g., weather forecasting \cite{weather}, water flow regulation \cite{water}, fire detection, diffusion of pollutants \cite{pollution}, smart grids \cite{smart-grids}, vehicular traffic \cite{traffic}.
	The problem of fusing data from different sensors can be accomplished either in a \textit{centralized} way, i.e. when there is a single fusion center collecting data from all sensors and taking care of the overall spatial domain of interest, or in \textit{distributed} (decentralized) fashion with multiple intercommunicating fusion centers (nodes) each of which can only access part of the sensor data and take care of a sub-region of the overall domain.
	The decentralized approach is preferable in terms of scalability of computation with the problem size and will be, therefore, undertaken in this paper.
	
	Since spatially distributed processes are usually modeled as infinite-dimensional systems, governed by \textit{partial differential equations} (PDEs), distributed state estimation for such systems turns out to be a key issue to be addressed.
While a lot of work has dealt with distributed consensus-type filters for finite-dimensional, both linear  \cite{Olfati1,Xiao,Calafiore,automatica} and nonlinear \cite{spl},  systems as well as for multitarget tracking \cite{J-STSP}, considerably less attention has been devoted to the more difficult case of distributed-parameter systems. 
	
Recent work \cite{Khan08,Stank09,Farina10,zhang-moura-krogh,pasqualetti} has addressed the design of distributed state estimators/observers for large-scale systems formed by the sparse interconnection of many subsystems (compartments).
Such systems are possibly (but not necessarily) originated from spatial discretization of PDEs. 
In particular, \cite{Khan08} presents a fully scalable distributed Kalman filter based on a suitable spatial decomposition of a complex large-scale system as well as on appropriate observation fusion techniques among the local Kalman filters.
In \cite{Stank09}, non-scalable consensus-based multi-agent estimators are proposed wherein each agent aims to estimate the state of the whole large-scale system.
In \cite{Farina10}, a moving-horizon partition-based approach  is followed in order to estimate the state of a large-scale interconnected system and decentralization is achieved via suitable approximations of covariances.
Further, \cite{zhang-moura-krogh} deals  with dynamic field estimation by wireless sensor networks with special emphasis on sensor scheduling for trading off communication/energy efficiency versus estimation performance.
In \cite{pasqualetti}, design of distributed continuous-time observers for partitioned linear systems is addressed.
	
As for the specific case of distributed-parameter systems, interesting contributions have been provided in \cite{Deme10,Deme13}
which present consensus filters wherein each node of the network aims to estimate the system state on the whole spatial domain of interest. 	

In the present paper, as compared to \cite{Deme10,Deme13}, a different strategy is adopted in which each node is only responsible for estimating the state over a sub-domain of the overall domain. 
This setup allows for a solution which is \textit{scalable} with respect to the spatial domain (i.e., the computational complexity in each node does not depend on the size of the whole spatial domain
	but only of its region of competence).  
	In this context, the contribution of the present paper is essentially in three directions.
	First, we develop \textit{scalable} consensus filters for distributed parameter systems by suitably adapting the so called Schwarz domain decomposition methods \cite{Schwa890,Lions88,Gander08,Chan94,Tos05,Smith96}, originally conceived to solve a boundary value problem
	by splitting it into smaller subproblems on subdomains and iterating to achieve consensus among the solutions on adjacent subdomains.
	Secondly, we exploit the \textit{finite element} (FE) method \cite{Pelosi09,Brenner96,Lee95} in order to approximate the original infinite-dimensional filtering problem into a, possibly large-scale, finite-dimensional one.
	Combining these two ingredients, we propose a novel distributed \textit{finite element Kalman filter} which generalizes to the more challenging distributed case previous work on FE Kalman filtering \cite{jap1,jap2}.
	Third, we provide results on the numerical stability of the proposed space-time discretization scheme as well as on the  stability of the proposed distributed FE Kalman filter.
Preliminary ideas on the topic can be found in \cite{ECC2015}.
	
	The rest of the paper is structured as follows. Section II introduces the basic notation and problem formulation.
	Then Section III  presents the centralized FE Kalman filter for distributed-parameter systems.
	Section IV shows how to extend such a filter to the distributed setting by means of parallel Schwarz consensus and analyzes the numerical stability in terms of boundedness and convergence of the discretization errors.
	Then, section V provides results on the exponential stability of the proposed distributed FE Kalman filter
	while section VI demonstrates its effectiveness  via numerical examples related to the estimation of a bi-dimensional temperature field.
	Finally, section VII ends the paper with concluding remarks and perspectives for future work.

	\section{Problem Formulation}
	This paper addresses the estimation of a scalar, time-and-space-dependent, field from given discrete, in both time and space, measurements related to such a field
	provided by multiple sensors placed within the domain of interest.
	The scalar field to be estimated $x \left( \mb{p}, t \right)$ is defined over the space-time domain $\Omega \times \rr_+$, as the solution of a
	\textit{partial differential equation (PDE)} of the form
	\be
	\dfrac{\partial x}{\partial t} +  \mathcal{A}(x) ~=~ f
	\label{PDE}
	\ee
	with (possibly unknown) initial condition $x \left( \mb{p}, 0 \right) = x_0(\mb{p})$, $\mb{p} \in \Omega$, and homogeneous boundary conditions
	\be
	\mathcal B (x) ~=~ 0 \mbox{ on } \partial \Omega\, .
	\label{boundary}
	\ee
	The space domain $\Omega$ is supposed to be bounded and with smooth boundary $ \partial \Omega$.
	
	The measurements
	\be
	y_{q,i} ~=~ h_i \left( x \left( \mb{s}_i, t_q \right) \right) + v_{q,i}
	\label{meas}
	\ee
	are provided by sensors $i \in \mathcal{S} \defi \{ 1, \dots, S \}$, located at positions $\mb{s}_i \in \Omega$, at discrete sampling instants $t_q$, $q \in
	\Z_+ = \{ 1, 2, \dots \}$, such that $0 < t_1 < t_2 < \cdots$.
	In (\ref{PDE})-(\ref{meas}): $\mb{p} \in \Omega$ denotes the $d$-dimensional ($d \in \{ 1, 2, 3 \}$) position vector; 
	$\mathcal{A}(\cdot)$ and $\mathcal{B}(\cdot)$ are linear operators over a suitable Hilbert space $V$, with $\mathcal A(\cdot)$ self-adjoint; 
	$f \left( \mb{p}, t \right)$ 
	is a forcing term possibly affected by process noise; 
	$h_i(\cdot)$ is the measurement function of sensor $i$; $v_{q,1}, \dots, v_{q,N}$ are mutually independent white measurement noise sequences, also independent from the initial state 
	$x_0(\mb{p})=x \left( \mb{p}, 0 \right)$ for any $\mb{p} \in \Omega$.
	
	More precisely, the aim is to estimate $ x (\mb{p},t ) $ given the information set $Y^t \defi \left\{ y_{q,i}, \forall i \in \mathcal{S} ~\mbox{and}~
	\forall q: t_q \leq t \right\}$.
	This is clearly an infinite-dimensional filtering problem.
	In the next section, it will be shown how it can be approximated into a finite-dimensional filtering problem by exploiting the FE
	method \cite{Pelosi09}-\cite{Brenner96}.
	
	An example of the above general problem is the estimation of the temperature field $x$ over the spatial domain of 
	interest given point measurements of temperature sensors. In this case, $V$ is usually taken as the Sobolev space $H^1(\Omega)$, the measurement function is simply $h(x)=x$, while
	the PDE (\ref{PDE}) reduces to the well known \textit{heat equation} with
	$\mathcal{A}(x) = -  \nabla \cdot \left ( \lambda \nabla (x) \right )$ and  $\mathcal B (x) = \alpha \,  {\partial x}/{\partial \mb{n}} + \beta x$ 
	with $\alpha(\mb{p}) \beta(\mb{p}) \ge 0 $, $\alpha(\mb{p}) + \beta(\mb{p}) > 0 $, $ \forall \mb{p} \in \partial \Omega$.
	Here $\lambda(\mb{p})$ is the \textit{thermal  diffusivity}, $\cdot$ stands for scalar product,
	$\nabla  \defi \partial  / \partial \mb{p} $ denotes the gradient operator, $\mb{n}$ is the outward pointing unit normal vector of the boundary 
	$\partial \Omega$, and $ {\partial x}/{\partial \mb{n}} = \nabla x \cdot \mb{n}$. Clearly, when the thermal  diffusivity is space-independent,
	one has $\mathcal{A}(x) = - \lambda \nabla^2 (x) $, where $\nabla^2 = \nabla \cdot \nabla$ is the Laplacian operator.
	
	Notice that considering homogeneous boundary conditions as in (\ref{boundary}) is not restrictive, since the non-homogeneous case $\mathcal B (x) = g$ on $\partial \Omega$
	can be subsumed into the  homogeneous one by means of the change of variables $z = x-w$, where $w$ is any function belonging to $V$ and satisfying the non-homogeneous boundary conditions.
	
	\section{Centralized Finite Element Kalman Filter}
	In this section, it is shown how to approximate the continuous-time infinite-dimensional system (\ref{PDE}) into a
	discrete-time finite-dimensional linear dynamical system within the FE framework. 
	
	By subdividing the domain $\Omega$ into a suitable set of non overlapping regions, or
	elements, and by defining a suitable set of basis functions $\phi_{j} (\mb{p}) \in V \, (j=1,\ldots,n)$ 
	on them, it is possible to write an approximation of the unknown function $x(\mb{p},t)$ as
	\be
	x(\mb{p},t) \approx \sum_{j=1}^{n} \phi_{j}(\mb{p}) \, x_j(t) ~=~ \boldsymbol{\phi}^T(\mb{p}) \, \mb{x}(t)
	\label{EXPA}
	\ee
	where: $x_j(t)$ is the unknown expansion coefficient of function $x(\mb{p},t)$ relative to time $t$ and 
	basis function $\phi_j(\mb{p})$; $\bs{\phi}(\mb{p})  \defi col \{ \phi_j (\mb{p} ) \}_{j=1}^n$ and $\mb{x}(t) \defi col \{ x_j(t) \}_{j=1}^n$.
	
	The choices of the basis functions $\phi_{j}$ and of the elements are key points of the FE method. Typically,
	the elements (triangles or quadrilaterals in 2D, tetrahedral or polyhedral in 3D) define a FE mesh with vertices $\mb{p}_j \in \Omega, j=1,\ldots,n $.
	Then each basis function $\phi_{j}$ is a piece-wise polynomial which vanishes outside the FEs around $\mb{p}_j$ and 
	such that $\phi_{j}(\mb{p}_i) = \delta_{ij}$, $\delta_{ij}$ denoting the Kronecker delta.
	
	In order to apply the Galerkin weighted residual method, let the PDE (\ref{PDE}) be recast in the  following (weak) integral form
	\begin{equation}
	\ba{l}
	\displaystyle{\int_\Omega} \dfrac{\partial x}{\partial t}  \, \psi \, d\mb{p}  + \displaystyle{\int_\Omega} \mathcal A(x) \, \psi \,  d\mb{p} = \displaystyle{\int_\Omega} f \, \psi d\mb{p} 
	\ea
	\label{PDE:weak}
	 \end{equation}
	 where $\psi (\mb{p})$ is a generic space-dependent weight function. The following assumption is now needed. \vspace{.3cm}
	 
	 \begin{enumerate}[\bf {A}1.]
	 \item Under the boundary conditions (\ref{boundary}), the quadratic form 
	 ${\int_\Omega} \mathcal A(\psi) \, \psi \,  d\mb{p} $ is bounded and coercive (i.e., positive definite).
	\end{enumerate}
	\vspace{.3cm}
	 
	 Then, by choosing the test function  $\psi (\mb{p})$ equal to the selected basis functions and exploiting the approximation (\ref{EXPA}) in (\ref{PDE:weak}),
	thanks to the linearity of operator $\mathcal{A}(\cdot)$ the 
	usual FE weak form is obtained \cite{Pelosi09}-\cite{Brenner96}
	\begin{equation}
	\ba{l}
	\underbrace{\left[ \displaystyle{\int_\Omega}  \bs{\phi}(\mb{p}) \bs{\phi}^T(\mb{p}) d\mb{p} \right]}_{\mb{M}} \dot{\mb{x}}(t) +
	\underbrace{\left[ \displaystyle{\int_\Omega} \bs{\phi}(\mb{p}) \left[ \mathcal{A} \left( \bs{\phi}(\mb{p}) \right) \right]^T d\mb{p} \right]}_{\mb{S}} \mb{x}(t) \vspace{2mm} \\
	=  \underbrace{\displaystyle{\int_\Omega} \bs{\phi}(\mb{p}) f(\mb{p},t) d\mb{p}}_{\mb{u}(t)}
	\ea
	\label{WF}
	\end{equation}
	where $\mathcal{A} \left( \bs{\phi} \right) \defi col \, \{ \mathcal{A}(\phi_j) \}_{j=1}^n$.
	It is evident how the first two integrals in (\ref{WF}) depend only on basis functions and can 
	be computed {\em a priori}. In particular, the first integral yields the well known 
	\textit{mass} matrix  $\mb{M}$, while the second depends on the operator $\mathcal{A}(\cdot)$ and, in the thermal case, 
	is the stiffness matrix $\mb{S}$ \cite{Pelosi09}. The third integral depends on the forcing term $f$, which is assumed to be known, and can hence
	be computed {\em a priori}, leading to a time dependent vector contribution $\mb{u}(t)$.
	
	It is worth pointing out that, in the FE weak form (\ref{WF}), the boundary conditions (\ref{boundary}) can be accounted for in two different ways \cite{Pelosi09,Brenner96}. 
	The so-called {\em essential} boundary conditions
	are handled by imposing them on the solution, i.e., by choosing basis functions belonging to $V_0 = \{x \in V: \; \mathcal B(x) = 0 \mbox{ on }  \partial \Omega\}$. On the other hand, the so-called {\em natural}
	boundary conditions can be directly incorporated into the weak form (\ref{PDE:weak}). 
	For example, in the case of the heat equation,
	the (isotherm) homogeneous Dirichlet boundary conditions $x = 0 \mbox{ on } \partial \Omega$ are essential, while the (adiabatic) homogeneous Neumann boundary conditions 
	$\partial x / \partial \mb{n} = 0$ are natural. Of course, by letting the functions $\alpha$ and $\beta$ vary on $\partial \Omega$, we can also have a problem with mixed essential/natural boundary conditions.
	In all the cases, the resulting linear differential equation is of the form
	\be \label{eq:diffeq}
	\mb{M} \, \dot{\mb{x}} + \mb{S} \, \mb{x} = \mb{u} + \bm{\epsilon}
	\ee
	where $\bm{\epsilon}$ arises from the approximation error\footnote{If $x$ is sufficiently smooth, then the FE approximation error is point-wise bounded and converges to zero as the size of the FE mesh tends to zero.}  in the finite-	dimensional representation (\ref{EXPA})  of $x$ in terms of basis functions. 
	Notice that $\mb{M}$ turns out to be positive definite by linear independence of the basis functions $\phi_j(\cdot)$. Further, $\mb{S}$ is positive definite as well thanks
	to the coercivity of the quadratic form in the left-hand side of (\ref{PDE:weak}).
	System (\ref{eq:diffeq}) can be discretized in time by different methods 
	(e.g., backward or forward Euler integration, or the zero-order-hold method)
	to provide the discrete-time state-space model
	\be
	\mb{x}_{k+1} = \mb{A}\mb{x}_{k} + \mb{B} \mb{u}_k + \mb{w}_k
	\label{DTM}
	\ee
	where the process noise $\mb{w}_k$ has been introduced to account for the various uncertainties and/or imprecisions (e.g. FE approximation, time discretization, and imprecise knowledge of boundary conditions).
	Specifically, the backward Euler method (here adopted for stability issues) leads to a marching in time FE implementation \cite{Lee95}
	which yields (\ref{DTM}) with
	$$
	\ba{l}
	\mb{A} = \left( \mb{I} + \Delta \mb{M}^{-1} \mb{S} \right)^{-1}, ~
	\mb{B} = \mb{A} \mb{M}^{-1} \Delta, \\
	\mb{u}_k \defi \mb{u}((k+1) \Delta), \mb{x}_k \defi \mb{x}(k \Delta) = col \{ x_j(k \Delta) \}_{j=1}^n
	\ea
	$$
	where $\Delta$ denotes the time integration interval. Notice that $\mb{A} $ is well defined for any $\Delta > 0$ since both $\mb{M}$ and $\mb{S}$ are positive definite.
	
	In the following, for the sake of notational simplicity, it will be assumed that each sampling instant is a multiple of $\Delta$,
	i.e., $t_q = T_q \Delta$ with $T_q \in \mathbb Z_+$, and we let $\mathcal T =\{T_1, T_2, \ldots \}$; irregular sampling could, however, be easily dealt with. This amounts to assuming that the numerical integration
	rate of the PDE (\ref{PDE}) in the filter can be higher than the measurement collection rate, which can be useful in order to reduce  numerical errors.
	In a centralized setting where all sensor measurements are available to the filter, the measurement equation (\ref{meas}) takes the discrete-time form
	\be
	\mb{y}_k ~=~  \mb{h} \left( \mb{x}_k \right) + \mb{v}_k
	\label{DTmeas}
	\ee
	for any $k = T_q \in \mathcal T$,
	where
	$$
	\ba{l}
	\mb{y}_k \defi col \left\{ y_{q,i} \right\}_{i\in \mathcal{S}},  \, \mb{h} \left( \mb{x} \right) \defi col \left\{ h_i \left( \bs{\phi}^T(\mb{s}_i) \mb{x} \right) \right\}_{i \in \mathcal{S}}, \\
	\mb{v}_k \defi col \left\{ v_{q,i} \right\}_{i \in \mathcal{S}}
	\ea
	$$
	In particular, in the case wherein all sensors directly measure the target field $x$, i.e. $h_i(x) = x$ for all $i \in \mathcal{S}$, the measurement equation
	(\ref{DTmeas}) turns out to be linear with $\mb{h}(\mb{x})= \mb{C} \mb{x}$, where
	\be
	\mb{C} = col \left\{ \bs{\phi}^T(\mb{s}_i) \right\}_{i \in \mathcal{S}}
	\label{outm}
	\ee
	Summarizing, the original infinite-dimensional continuous-time problem has been reduced to a much simpler finite-dimensional (possibly large-scale) discrete time filtering problem
	(a linear one provided that all sensor measurement functions are linear) to which the \textit{Kalman filter}, or \textit{extended Kalman filter} when sensor
	nonlinearities are considered, can be readily applied.
	The resulting centralized filter recursion becomes:
	\begin{eqnarray}
	&& \hspace{-.8cm} \mb{\hat{x}}_{k|k} = \left \{
	\ba{ll}
	\mb{\hat{x}}_{k|k-1} + \mb{L}_k
	\left( \mb{y}_k - \mb{h} \left( \mb{\hat{x}}_{k|k-1} \right) \right) & \mbox{ if } k \in \mathcal T \\
	\mb{\hat{x}}_{k|k-1}  & \mbox{ otherwise }
	\ea \right . \nonumber \\
	&&  \hspace{-.8cm} \mb{P}_{k|k} = \left \{
	\ba{ll}
	\mb{P}_{k|k-1} - \mb{L}_k  
	\mb{C}_k^T \mb{P}_{k|k-1} \quad \quad \quad \quad & \mbox{ if } k \in \mathcal T \\
	\mb{P}_{k|k-1}   & \mbox{ otherwise }
	\ea \right . \nonumber \\
	&& \hspace{-.8cm} \mb{\hat{x}}_{k+1|k}  =  \mb{A} \mb{\hat{x}}_{k|k} + \mb{B} \mb{u}_k \nonumber \\
	&& \hspace{-.8cm} \mb{P}_{k+1|k}  =  \mb{A} \mb{P}_{k|k} \mb{A}^T + \mb{Q}_k \label{KF}
	\end{eqnarray}
	where
	\begin{eqnarray*}
		&& \mb{C}_k  =  \dfrac{\partial \mb{h}}{\partial \mb{x}} \left( \mb{\hat{x}}_{k|k-1} \right)  \\
		&& \mb{L}_k  =  \mb{P}_{k|k-1} \mb{C}_k \left( \mb{R}_k + \mb{C}_k \mb{P}_{k|k-1} \mb{C}_k^T \right)^{-1}
	\end{eqnarray*}
	for $k \in \mathcal T$. The recursion is initialized from suitable $\mb{\hat{x}}_{1|0}$ and $\mb{P}_{1|0} = \mb{P}_{1|0}^T > \mb{0}$.
	In (\ref{KF}), $\mb{Q}_k$ and $\mb{R}_k$ denote the covariance matrices of the process noise $\mb{w}_k$ and, respectively, measurement noise $\mb{v}_k$, which are assumed as usual to be white, zero-mean, mutually uncorrelated and also uncorrelated with the initial state $\mb{x}_1$.

	
	\section{Distributed Finite Element Kalman Filter}
	
	In order to develop a scalable distributed filter for monitoring the target field, the idea is to decompose the original problem on the whole domain
	of interest into estimation subproblems concerning smaller subdomains, and then to assign such subproblems to different \textit{nodes} which can locally process and exchange data.
	To this end, let us consider the set of nodes $\mathcal{N} = \{ 1, \dots, N \}$, subdivide the domain $\Omega$ into possibly overlapping subdomains $\Omega_m$, $m \in \mathcal{N}$, such that $\Omega = \bigcup_{m \in \mathcal{N}} \Omega_m$, 
	and assign the task ``estimation of $x$ over $\Omega_m$'' to node $m$.
	Further, let 
	$\mb{y}_q^m \defi col \left\{ y_{q,i} : \mb{s}_i \in \Omega_m \right\}$ denote the vector of local measurements available to node $m$ at time $t_q$.
	
	Hence, the idea is to run in each node $m \in \mathcal{N}$ a field estimator for the region $\Omega_m$ exploiting local measurements
	$\mb{y}_q^m$, information from the nodes assigned to neighboring subdomains, as well as the PDE model (\ref{PDE}) properly discretized in time and space.
	Taking inspiration from the Schwarz method \cite{Schwa890,Lions88,Gander08}, neighboring local estimators should iteratively find a \textit{consensus}
	on the estimates concerning the common parts. The Schwarz method  has been originally conceived \cite{Schwa890} for an iterative solution of boundary value problems.
	Subsequently, it has received renewed interest \cite{Lions88,Gander08} in connection with the parallelization of PDE solvers.
	In loose terms, the idea of the \textit{parallel Schwarz method} is to decompose the original PDE problem on the overall domain of interest into subproblems concerning smaller subdomains, and then to solve in parallel such subproblems via iterations in which previous solutions concerning neighboring 
	subdomains are used as boundary conditions.
	As shown below, such an idea turns out to be especially useful for the distributed filtering problem considered in this work.
	
	To formalize the consensus let us define, for any $m \in \mathcal{N}$, a partition $\left\{ \Gamma_{mj} \right\}_{j \in \mathcal{N}_m}$ of $\partial\Omega_m$ (the boundary of $\Omega_m$) such that
	\be
	\ba{l}
	\Gamma_{mm} = \partial \Omega \cap \partial \Omega_m \vspace{1mm} \\
	\partial \Omega_m = \displaystyle{\bigcup_{j \in \mathcal{N}_m}} \Gamma_{mj} \vspace{1mm} \\
	\Gamma_{mj} \subset \Omega_j ,~~~ \forall j \neq m \vspace{1mm} \\
	\Gamma_{mj} \cap \Gamma_{mh} = \emptyset,~~~ \forall j \neq h
	\ea
	\label{part}
	\ee
	In this way, each piece $\Gamma_{mj}$ of $\partial \Omega_m$ for any $j \in \mathcal{N}_m \backslash \{ m \}$ is uniquely assigned to node $j$. 
	Notice that in the above definitions,  for each node $m$, $\mathcal{N}_m$ indicates the in-neighborhood of node $m$, where $j$ is called an in-neighbor of node $m$ whenever $ \Gamma_{mj} \ne \emptyset$
	(by definition, $\mathcal{N}_m$ includes the node $m$.) This clearly originates a directed network (graph)
	$\mathcal{G} = \left( \mathcal{N}, \mathcal{L} \right)$
	with node set $\mathcal{N}$ and link set $\mathcal{L} \defi \{ (j,m) \in \mathcal{N} \times \mathcal{N}: \Gamma_{mj} \ne \emptyset \}$.
	\begin{figure}[h!]
		\centering
		\includegraphics[width=.7\columnwidth]{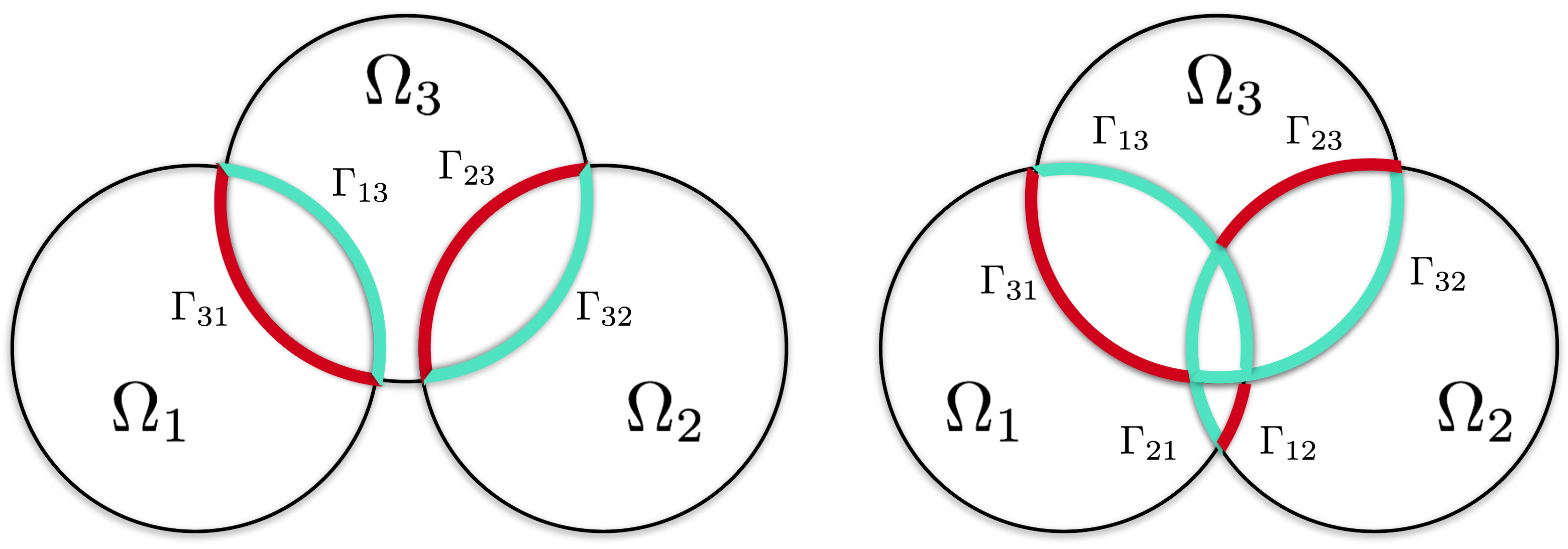}
		\caption{Definition of interfaces $\Gamma_{mj}$ in two different configurations with three overlapping subdomains.}
		\label{fig:interfaces}
	\end{figure}
	
	In order to describe the filtering cycle to be implemented in node $m$ within the sampling interval $[t_q,t_{q+1})$, let us assume that at time $t_q^-$, before the
	acquisition of $\mb{y}_q^m$, such a node is provided with a prior estimate ${\hat{x}}_{q|q-1}^m$ 
	as the result of the previous filtering cycles.
	Let $\delta$ be the time interval necessary for performing one consensus step, i.e., information exchange between neighbors and related computations. Then,
	$L_q \defi  \left( t_{q+1} - t_q \right) / \delta$ represents the number of consensus steps  (equal to the number of allowed data exchanges) in the $q$-th
	sampling interval. Note that, for the sake of notational simplicity, hereafter it is supposed that $ t_{q+1} - t_q$ is an integer multiple of $\delta$, i.e., $L_q \in \mathbb Z_+$.
	Anyway, the method could easily encompass the general case.
	Then, the above mentioned filtering cycle for the proposed distributed estimation algorithm essentially consists of:
	\begin{enumerate}
		\item \textbf{Correction}, i.e. incorporation (assimilation) of the last measurement $\mb{y}_q^m$ into the current estimate;
		\item \textbf{Consensus}, i.e. alternate exchanges of estimates with the neighborhood $\mathcal{N}_m$ and predictions over the time sub-intervals
		$[t_q + (\ell-1) \delta, t_{q} + \ell \delta]$ for  $\ell = 1, \dots, L_q$, i.e. $L_q$ times.
	\end{enumerate}
	The proposed \textit{Parallel Schwarz Consensus} filter is detailed hereafter. \vspace{.3cm}
	
	{\bf Algorithm 1.}
	\begin{enumerate}
		\item Given $\mb{y}_q^m$, update the prior estimate ${\hat{x}}_{q|q-1}^m$ 
		into ${\hat{x}}_{q|q}^m$. 
		\item Initialize the consensus with ${\hat{x}}_{q,0}^m = {\hat{x}}_{q|q}^m$ 
		and ${\hat{x}}_{q,-1}^m ={\hat{x}}_{q|q}^m$. 
		\item For $\ell = 1, \dots, L_q$ proceed as follows
		\begin{enumerate}
			\item Exchange data with the neighborhood; specifically send to neighbor $j$ the data ${\hat{x}}_{q,\ell-1}^m$ 
			concerning the sub-boundary $\Gamma_{jm} \subset \partial \Omega_j$ , and
			get from neighbor $j$ the data ${\hat{x}}_{q,\ell-1}^j$ 
			concerning the sub-boundary $\Gamma_{mj} \subset \partial \Omega_m$.
			\item Solve the problem
			\be
			\dfrac{\hat{x}^m_{q,\ell} - \hat{x}_{q,\ell-1}^m}{\delta} + \mathcal{A} \left( \hat{x}_{q,\ell}^m \right) ~=~ f_{q,\ell}  ~~~\mbox{in} ~ \Omega_m
			\label{PDE2}
			\ee
			subject to the Dirichlet boundary conditions
			\be
			\hat{x}_{q,\ell}^m = \hat{x}^j_{q,\ell-1} ~~~\mbox{on}~ \Gamma_{mj} ~~~~ \forall j \in \mathcal{N}_m \backslash \{m \}
			\label{DBC}
			\ee
			and the linear boundary conditions
			\be
			\mathcal B (\hat{x}_{q,\ell}^m) = 0 \mbox{ on } \Gamma_{mm} \, .
			\label{BC}
			\ee
			where $f_{q,\ell}(\mb{p}) \defi f \left( \mb{p}, t_q + \ell \delta \right)$.
		\end{enumerate}
		\item Set ${\hat{x}}_{q+1|q}^m = {\hat{x}}_{q,L_q}^m$  
		for the next cycle.
	\end{enumerate}
	\vspace{.3cm}
	
	Some remarks concerning the above reported algorithm are in order.  As it can be seen from step 3b), the information received
	by neighboring nodes is taken into account by explicitly imposing the non-homogeneous Dirichlet interface conditions (\ref{DBC}) on $\Gamma_{mj}, j \in \mathcal{N}_m \setminus \{m\}$.
	Clearly, a delay is introduced in those terms concerning neighboring nodes which makes the algorithm well-suited for distributed computation.
	With this respect, it is worth pointing out that the proposed consensus algorithm is based on the parallel Schwarz method for evolution problems,
	which, as well known, enjoys nice convergence properties to the centralized solution as the time discretization step $\delta$ tends to zero \cite{Lions88}-\cite{Gander08}.
	Hence, it seems a sensible and promising approach to spread the information through the network. 
	Finally, notice that the prediction step of each local filter is directly incorporated into the consensus algorithm.

	\subsection{Implementation via the finite-element method}
	
	In practice, the algorithm, and in particular the solution of the boundary value problem (\ref{PDE2})-(\ref{BC}), has to be implemented
	via a finite dimensional approximation. In particular, we follow the same approach described in Section III for the centralized case
	by constructing a FE mesh for the global domain $\Omega$, and then decomposing such a grid into $N$ overlapping sub-meshes, according to the domain decomposition. 
	For the sequel, it is important to distinguish vertices lying on the boundary between neighbors (\textit{interface}) from the other vertices of the subdomain. 
	To this end, let ${\rm int} (S)$ denote the interior of a generic set $S$. 
	Then, we introduce the sets of indices $\mathfrak{I}_m \defi \{ i: \mb{p}_i \in {\rm int} (\Omega_m) \cup \Gamma_{mm} \}$ 
	and $\mathfrak{I}_{mj} \defi \{ i: \mb{p}_i \in \Gamma_{mj} \}$
	of the basis functions corresponding to \textit{internal} and, respectively, \textit{interface} vertices of subdomain $\Omega_m$.
	In particular, let 
	$\mb{x}^m \defi col \{ x_i : i \in \mathfrak{I}_m \}, \, m = 1,\dots, N$,
	denote the vector of field values in vertices belonging to $ {\rm int} (\Omega_m) \cup \Gamma_{mm} $,
	i.e. the \textit{internal} state of subsystem $m$.
	Then, it is possible to extract from (\ref{eq:diffeq}) the rows relative to states $\mb{x}^m$ so that
	\begin{eqnarray}
	&\mb{M}^{mm}& \dot{\mb{x}}^m + \sum_{j \in \mathcal N_m \setminus \{m\}} \mb{M}^{m j} \dot{\mb{x}}^j 
	+ \mb{S}^{mm} \mb{x}^m \nonumber \\ && {}  + \sum_{j \in \mathcal N_m \setminus \{m\}} \mb{S}^{m j} \mb{x}^j =   
	\mb{u}^m + \bm{\epsilon}^m
	\label{localaugsys}
	\end{eqnarray}
	where the matrices $\mb{M}^{mj}$ and $\mb{S}^{mj}$ take into account the contribution of state variables in vertices $\mb{p}_j \in \Gamma_{mj}$, and $\bm{\epsilon}^m$ accounts for the approximation error in the finite-dimensional 
	representation \eqref{EXPA} of $x$ in terms of basis functions.	
	Notice that both $\mb{M}^{mm}$ and $\mb{S}^{mm}$ are positive definite because so are $\mb{M}$ and $\mb{S}$.
	As a result, the ODE (\ref{eq:diffeq}) can be written 
	as the interconnection of $N$ subsystems of the form (\ref{localaugsys}).	 
		
	Each of the subsystems (\ref{localaugsys}) can be discretized in time in the interval $[t_q,t_{q+1}]$
	using a modified backward Euler technique wherein a delay is introduced in those terms concerning neighboring nodes, so that at time $t_q + \ell \delta$ 
	we obtain the following discrete-time linear descriptor system 
	\begin{eqnarray}
	&&{\mb{M}}^{mm} \left(\frac{{\mb{x}}^m_{q,\ell+1}-{\mb{x}}^m_{q,\ell}}{\delta}\right) 
	+ {\mb{S}}^{mm} \, {\mb{x}}^m_{q,\ell+1} \nonumber \\ && {} 
	+  \sum_{j \in \mathcal N_m \setminus \{m\}}   \left[ {\mb{M}}^{mj} \, \left(\frac{{\mb{x}}^j_{q,\ell}-{\mb{x}}^j_{q,\ell-1}}{\delta}\right) + {\mb{S}}^{mj} \, {\mb{x}}_{q,\ell}^j  \right ]\nonumber \\ && {} 
	=  
	{\mb{u}}^m_{q,\ell+1} + {\bm{\epsilon}}^m_{q,\ell+1} + \bm{\tau}^m_{q,\ell}
	\label{time-discr:m}
	\end{eqnarray}
	where ${\mb{x}}^m_{q,\ell} \defi {\mb{x}}^m (t_q + \ell \delta)$, for $\ell = 1 \, \ldots, L$, and $\bm{\tau}_{q,\ell}^m$ denotes the time discretization error at time $t_q + \ell \delta$. The recursion (\ref{time-discr:m}) is 
	initialized at time $t_q$ by setting
	\begin{equation}
	\ba{l}
	{\mb{x}}^m_{q,0} =  {\mb{x}}^m (t_q) , 
	\\ {\mb{x}}^j_{q,0} =  {\mb{x}}^j (t_q), \quad {\mb{x}}^j_{q,-1} =  {\mb{x}}^j (t_q) , \quad j \in \mathcal N_m \setminus \{m\}
	\ea
	\label{eq:init}
	\end{equation}
	The well-posedness of the discretization scheme resulting from (\ref{time-discr:m})-(\ref{eq:init}) will be analyzed in Section IV-B.
	
	It can be readily seen that such a hybrid Euler time discretization implements 
	the Parallel Schwarz method, described earlier. In fact, it is equivalent to approximate $x$ in $\Omega_m$ at time $t_q + \ell \delta$ as
		\begin{eqnarray}
		x(\mb{p}, t_q + \ell \delta) &\approx& \displaystyle \sum_{i \in \mathfrak{I}_m} \phi_{i}^m(\mb{p}) \, x_{q,\ell}^{m,i}     \\ && {} 
		+ \displaystyle \sum_{j \in \mathcal{N}_m \setminus \{m\}}  \displaystyle \sum_{i \in \mathfrak{I}_{mj}} \phi_{i}^j(\mb{p}) \, x_{q,\ell-1}^{j,i} \nonumber
		\label{approx_Dirich}
		\end{eqnarray}
	which in turn corresponds to explicitly imposing non-homogeneous Dirichlet interface conditions on $\Gamma_{mj}, j \in \mathcal{N}_m \setminus \{m\}$ taken from neighboring nodes (like in  (\ref{DBC})).

	Thanks to the positive definiteness of ${\mb{M}}^{mm}$ and ${\mb{S}}^{mm}$, each discretized-model (\ref{time-discr:m}) can be easily transformed into a state-space model of the form
	\begin{eqnarray} \label{statespace_local}
	{\mb{x}}^m_{q,\ell} &=& \mb{A}^m {\mb{x}}^m_{q,\ell-1} + \sum_{j \in \mathcal N_m \setminus \{m\}} \mb{A}^{m j} \hat{\mb{x}}^j_{q,\ell-1}      \\ && {} 
	+ \sum_{j \in \mathcal N_m \setminus \{m\}} \bar{\mb{A}}^{m j} {\mb{x}}^j_{q,\ell-2}
	+ \mb{B}^{m} \mb{u}_{q,\ell}^m + {\mb{w}}_{q,\ell}^m  \nonumber
	\end{eqnarray}
	where
	\begin{equation}
	\begin{aligned}
	\mb{A}^m & = \left( \mb{M}^{mm} + \delta \mb{S}^{mm} \right)^{-1} \mb{M}^{mm} \\
	\mb{A}^{m j} &= \left( \mb{M}^{mm} + \delta \mb{S}^{mm} \right)^{-1} \left( - \delta \mb{S}^{mj} - \mb{M}^{mj} \right)  \\
	\bar{\mb{A}}^{m j} &= \left( \mb{M}^{mm} + \delta \mb{S}^{mm} \right)^{-1} \mb{M}^{mj}     \\
	\mb{B}^m &= \left( \mb{M}^{mm} + \delta \mb{S}^{mm} \right)^{-1} \delta 
	\end{aligned} \nonumber
	\end{equation}
	and 
	${\mb{w}}_{q,\ell}^m = ({\mb{M}}^{mm} + \delta  \, {\mb{S}}^{mm} )^{-1} \delta \left( 
	\tilde{\bm{\epsilon}}_{q,\ell+1}^m + \bm{\tau}_{q,\ell}^m \right)$ 
	is the error combining the effects of both spatial and temporal discretizations. 

	Such interconnected models can be exploited so as to derive a FE approximation of the distributed-state estimation algorithm with Parallel Schwarz Consensus (Algorithm 1). In particular, 
	the numerical solution of (\ref{PDE2})-(\ref{BC}) takes the form of the local one-step-ahead predictor for model (\ref{statespace_local}) at time $t_q + (\ell-1) \delta$, 
	whereas the correction step of the local filtering cycle is the usual (extended) Kalman filter update step for the local
	subsystem. The resulting distributed finite-element (extended) Kalman filter is as follows. \vspace{.3cm}
	
	{\bf Algorithm 2.}
		\begin{enumerate}
		\item Given $\mb{y}_q^m$, update the prior estimate $\mb{\hat{x}}_{q|q-1}^m$ and covariance $\mb{P}_{q|q-1}^m$ into
		$\mb{\hat{x}}_{q|q}^m$ and $\mb{P}_{q|q}^m$ as follows
			\begin{eqnarray}
	\mb{\hat{x}}_{q|q}^m &=& \mb{\hat{x}}^m_{q|q-1} + \mb{L}_q^m
	\left( \mb{y}_q^m - \mb{h}^m \left( \mb{\hat{x}}^m_{q|q-1} \right) \right)
	\nonumber \\
	\mb{P}_{q|q}^m &=& \mb{P}_{q|q-1}^m - \mb{L}_q^m  
	(\mb{C}_q^m)^T \mb{P}^m_{q|q-1}  \nonumber \\
	\mb{C}_q^m  &=&  \dfrac{\partial \mb{h}^m}{\partial \mb{x}} \left( \mb{\hat{x}}^m_{q|q-1} \right) \nonumber  \\
	\mb{L}_q^m  &=&  \mb{P}_{q|q-1}^m \mb{C}_q^m \left( \mb{R}_q^m + \mb{C}_q^m \mb{P}_{q|q-1}^m (\mb{C}_q^m)^T \right)^{-1} \nonumber 
	\end{eqnarray}
	where $\mb{h}^m \defi col \left\{ h_{i} : \mb{s}_i \in \Omega_m \right\}$ denote the local measurement function at node $m$.
		
		\item Initialize the consensus with $\mb{\hat{x}}_{q,0}^m =\mb{\hat{x}}_{q|q}^m$, $\mb{P}_{q,0}^m =\mb{P}_{q|q}^m$ and $\mb{\hat{x}}_{q,-1}^m =\mb{\hat{x}}_{q|q}^m$, $\mb{P}_{q,-1}^m =\mb{P}_{q|q}^m$.
		\item For $\ell = 1, \dots, L_q$ proceed as follows
		\begin{enumerate}
			\item Exchange data with the neighborhood; specifically send to neighbor $j$ the data $\mb{\hat{x}}_{q,\ell-1}^m, \mb{P}^m_{q,\ell-1}$ concerning the 
			sub-boundary $\Gamma_{jm} \subset \partial \Omega_j$ , and
			get from neighbor $j$ the data $\mb{\hat{x}}_{q,\ell-1}^j, \mb{P}^j_{q,\ell-1}$ concerning the sub-boundary $\Gamma_{mj} \subset \partial \Omega_m$.
			\item set
			\begin{eqnarray}   \nonumber 
			\hat{\mb{x}}^m_{q,\ell} &=& \mb{A}^m \hat{\mb{x}}^m_{q,\ell-1} + \sum_{j \in \mathcal N_m \setminus \{m\}} \mb{A}^{m j} \hat{\mb{x}}^j_{q,\ell-1}      \\ && {}  \hspace{-1cm}
			+ \sum_{j \in \mathcal N_m \setminus \{m\}} \bar{\mb{A}}^{m j} \hat{\mb{x}}^j_{q,\ell-2}
			+ \mb{B}^{m} \mb{u}_{q,\ell}^m \label{statespace_local:pred} \\ 
		\mb{P}_{q,\ell}^m &=& \gamma^2 \, \mb{A}^m \mb{P}_{q,\ell-1}^m \left ( \mb{A}^m  \right )^T + \mb{Q}^m 
		\label{LDTM-cov}
		\end{eqnarray}
		with $\gamma \ge 1$.
		\end{enumerate}
		\item Set $\mb{\hat{x}}_{q+1|q}^m =\mb{\hat{x}}_{q,L_q}^m$  and $\mb{P}_{q+1|q}^m =\mb{P}_{q,L_q}^m$ for the next cycle.
	\end{enumerate}
	\vspace{.3cm}

	As previously shown, the additional terms $\sum_{j \in \mathcal N_m \setminus \{m\}} \mb{A}^{m j} \hat{\mb{x}}^j_{q,\ell-1}$ and $\sum_{j \in \mathcal N_m \setminus \{m\}} \bar{\mb{A}}^{m j} \hat{\mb{x}}^j_{q,\ell-2}$ 
	in equation (\ref{statespace_local}) arise from the non-homogeneous Dirichlet boundary conditions (\ref{DBC}). In this respect, it is worth noting that the matrices $\mb{A}^{m j}$ and $\bar{\mb{A}}^{m j}$ are 
	sparse since only the components of the neighbor estimates $\mb{\hat  x}^j_{q,\ell-1}$ and $\mb{\hat  x}^j_{q,\ell-2}$ concerning the sub-boundary $\Gamma_{mj}$ are involved.
	The positive real $\gamma > 1 $ is a covariance boosting factor whose role, as will be discussed in the stability analysis of the distributed FE-KF, is that of guaranteeing convergence of the estimates. 	
	The covariance boosting factor is also necessary in order to compensate for the additional uncertainty associated with the boundary conditions at the interfaces, i.e., for the uncertainty associated 
	with the estimates $\sum_{j \in \mathcal N_m \setminus \{m\}} \mb{A}^{m j} \hat{\mb{x}}^j_{q,\ell-1}$ and $\sum_{j \in \mathcal N_m \setminus \{m\}} \bar{\mb{A}}^{m j} \hat{\mb{x}}^j_{q,\ell-2}$ .
	In fact, such an uncertainty is not explicitly accounted for in (\ref{LDTM-cov}) due to the fact that the correlation between the estimates of neighboring nodes is not precisely known.
	The interested reader is referred to  \cite{Farina10} for additional insights on this issue  in the context of distributed estimation of 
	large-scale interconnected systems. As in the centralized context, the positive definite matrix $ \mb{Q}^m$ accounts for the various uncertainties and imprecisions (i.e., discretization errors, imprecise knowledge of the exogenous
	input $f$ and of the boundary conditions (\ref{BC})).


	\subsection{Numerical stability}
	
	As previously shown, in the FE-based implementation the Parallel Schwarz consensus amounts to performing a hybrid Euler discretization on the interconnection of the $N$ subsystems
	(\ref{localaugsys}). Hence, as a preliminary analysis step, it is important to verify the well-posedness of such a modified discretization method in terms of numerical stability (i.e., in terms of boundedness and convergence 
	of the time-discretization errors). To this end, it is convenient to consider the global dynamics of the interconnection.
	
	Let us consider the augmented global state $\mb{\tilde{x}} \defi col \{ \mb{x}^m, \, m = 1,\dots, N \}$, which clearly contains repeated components of the state due to the overlapping nature of the decomposition.
	Let the vectors $\tilde{\mb{u}}  $ and $\tilde{\bm{\epsilon}}$ be defined in a similar way.
	In terms of $\mb{\tilde{x}}$ the interconnection of the $N$ subsystems of the form (\ref{localaugsys}) gives rise to a global augmented system which obeys the following continuous-time 	linear dynamics 
	\be \label{eq:diffeqtilde}
	\tilde{\mb{M}} \, \dot{ \tilde{\mb{x}} } + \tilde{\mb{S}} \, \tilde{\mb{x}} = \tilde{\mb{u}} + \tilde{\bm{\epsilon}}
	\ee
	Note that the only difference between (\ref{eq:diffeq}) and (\ref{eq:diffeqtilde}) is the presence of duplicated states in the latter linear ODE. Nevertheless, the two systems originate an identical state evolution.
	According to the divide-and-conquer strategy, matrices $\tilde{\mb{M}}$ and $\tilde{\mb{S}}$ can be decomposed as
	\begin{eqnarray} \label{Mdecomp}
	\tilde{\mb{M}}  &=&  \tilde{\mb{M}}_D + \tilde{\mb{M}}_F 
	\\ \tilde{\mb{S}}  &=&  \tilde{\mb{S}}_D + \tilde{\mb{S}}_F \label{Sdecomp}
	\end{eqnarray} 
	with $ \tilde{\mb{M}}_D =$ block-diag($\mb{M}^{11},\dots,\mb{M}^{NN}$), $\tilde{\mb{S}}_D =$ block-diag($\mb{S}^{11},\dots,\mb{S}^{NN}$), whereas $\tilde{\mb{M}}_F$ and $\tilde{\mb{S}}_F$ take into account
	the FE interconnection structure among neighboring subsystems.
	By substituting (\ref{Mdecomp})-(\ref{Sdecomp}) into (\ref{eq:diffeqtilde}), one obtains
	\be \label{eq:diffeqtilde2}
	\tilde{\mb{M}}_D \, \dot{ \tilde{\mb{x}} } + \tilde{\mb{S}}_D \, \tilde{\mb{x}} + \tilde{\mb{M}}_F \, \dot{ \tilde{\mb{x}} } + \tilde{\mb{S}}_F \, \tilde{\mb{x}} = \tilde{\mb{u}} + \tilde{\bm{\epsilon}} \, .
	\ee
	Then, by applying the hybrid Euler time discretization (\ref{time-discr:m}), the time-discretized augmented system takes the form
	\begin{eqnarray}
	&&\hspace{-.5cm} \tilde{\mb{M}}_D \left(\frac{\tilde{\mb{x}}_{q,\ell+1}-\tilde{\mb{x}}_{q,\ell}}{\delta}\right) 
	+ \tilde{\mb{S}}_D\tilde{\mb{x}}_{q,\ell+1} 
	+ \tilde{\mb{M}}_F \left(\frac{\tilde{\mb{x}}_{q,\ell}-\tilde{\mb{x}}_{q,\ell-1}}{\delta}\right) \nonumber \\ && {} 
	+ \tilde{\mb{S}}_F\tilde{\mb{x}}_{q,\ell} =  
	\tilde{\mb{u}}_{q,\ell+1} + \tilde{\bm{\epsilon}}_{q,\ell+1} + \bm{\tau}_{q,\ell}
	\label{time-discr}
	\end{eqnarray}
	 for $\ell = 0, \ldots, L-1$, where $\tilde{\mb{x}}_{q,\ell} \defi \tilde{\mb{x}} (t_q + \ell \delta)$, and, as previously, $\bm{\tau}_{q,\ell}$ denotes the time discretization error at time $t_q + \ell \delta$. Further, the initialization
	(\ref{eq:init}) can be simply rewritten as
	\be
	\ba{l}
	\tilde{\mb{x}}_{q,0} = \tilde{\mb{x}} (t_q) \\
	\tilde{\mb{x}}_{q,-1} = \tilde{\mb{x}} (t_q) 
	\ea
	\label{eq:init:tilde}
	\ee
	
	The following result can now be stated which summarizes the numerical stability properties\footnote{The interested reader is referred to chapter 12 of \cite{num:stab} 
	for an introduction on the concepts of consistency, zero-stability, and convergence of time-discretization methods.} of (\ref{time-discr})-(\ref{eq:init:tilde}). \vspace{.3 cm}
	
	\begin{theorem} 
	The  hybrid Euler time-discretization scheme (\ref{time-discr})-(\ref{eq:init:tilde}) is {\em consistent} with local truncation error of order $1$. Further, it is {\em zero-stable} provided that the following condition holds
	\be\label{eq:rho}
	\rho ( \tilde{\mb{M}}_D^{-1} \, \tilde{\mb{M}}_F) < 1
	\ee
	where $\rho (\cdot)$ denotes the spectral radius.
	\end{theorem} \vspace{.3cm}
	{\em Proof:} Let $\mathcal{D}$ denote the differential operator in the left-hand side of (\ref{eq:diffeqtilde2}), i.e., 
	\[ 
	\mathcal{D} (\bm{\xi},t) = \tilde{\mb{M}}_D \, \dot{\bm{\xi}}(t)  +  \tilde{\mb{S}}_D \, \bm{\xi} (t) 
	+ \tilde{\mb{M}}_F \, \dot{\bm{\xi}}(t) +  \tilde{\mb{S}}_F \, \bm{\xi} (t)  
	\] 
	for any smooth time-function $\bm{\xi}$. Further, let $\mathcal{D}_\delta$ denote the discrete-time operator in the left-hand side of
	(\ref{time-discr}), i.e.,  
	\begin{eqnarray*}
	\mathcal{D}_\delta (\bm{\xi},t) &=&  \tilde{\mb{M}}_D   \left(\frac{\bm{\xi} (t+\delta) -  \bm{\xi} (t)}{\delta} \right ) +  \tilde{\mb{S}}_D \, \bm{\xi} (t+\delta) \\
	&& {} +  \tilde{\mb{M}}_F   \left(\frac{\bm{\xi} (t) -  \bm{\xi} (t-\delta)}{\delta} \right ) + \tilde{\mb{S}}_F \, \bm{\xi} (t) \, . 
	\end{eqnarray*}
	As well known, the time-discretization scheme (\ref{time-discr})  is consistent
	when, for any smooth time-function $\bm{\xi}$ and for any time $t$,  $\mathcal{D}_\delta (\bm{\xi},t)$ converges to $\mathcal{D} (\bm{\xi},t)$ as $\delta$ goes to $0$. By taking the Taylor expansion
	of $\bm{\xi}$ in $t$, we can write $ \bm{\xi} (t+\delta) = \bm{\xi} (t) + \delta \,  \dot{\bm{\xi}}(t)  + \delta^2 \,  \ddot{\bm{\xi}}(t)  + O(\delta^3)$ and $ \bm{\xi} (t-\delta) = \bm{\xi} (t) - \delta \,  \dot{\bm{\xi}}(t) 
	 + \delta^2 \,  \ddot{\bm{\xi}}(t)   + O(\delta^3)$.
	Hence, after some algebra, we have
	\begin{eqnarray*}
	\mathcal{D}_\delta (\bm{\xi},t) &=& \mathcal{D} (\bm{\xi},t) + \tilde{\mb{M}}_D \, \delta \, \ddot{\bm{\xi}}(t)  + \tilde{\mb{S}}_D \, \delta \, \dot{\bm{\xi}}(t) - \tilde{\mb{M}}_F \, \delta \, \ddot{\bm{\xi}}(t) \\
	&& {} + O (\delta^2)
	\end{eqnarray*}
	which shows that the scheme is consistent and the local truncation error has order $1$.
	
	In order to study zero-stability, we start by considering the limit for $\delta$ going to zero of the time-difference equation (\ref{time-discr}), which is given by
	\be\label{eq:zero:stability}
	\tilde{\mb{M}}_D  \left({\tilde{\mb{x}}_{q,\ell+1}-\tilde{\mb{x}}_{q,\ell}}{}\right) 
	+ \tilde{\mb{M}}_F \left({\tilde{\mb{x}}_{q,\ell}-\tilde{\mb{x}}_{q,\ell-1}}{}\right)  = 0 \, .
	\ee
	In fact, zero-stability of the time-discretization scheme (\ref{time-discr}) corresponds to the asymptotic stability of the discrete-time system (\ref{eq:zero:stability}).
	Then the proof can be concluded by noting that, by defining $\bm{\zeta}_{q,\ell+1}= \tilde{\mb{x}}_{q,\ell+1}-\tilde{\mb{x}}_{q,\ell}$, system  (\ref{eq:zero:stability}) can be rewritten as
	\[
	\left [ \begin{array}{c}
	\tilde{\mb{x}}_{q,\ell+1} \\ \bm{\zeta}_{q,\ell+1} 
	\end{array}
	\right ]
	=
	\left [ \begin{array}{cc}
	\mb{I} & - \tilde{\mb{M}}_D^{-1} \, \tilde{\mb{M}}_F\\
	\mb{0} & - \tilde{\mb{M}}_D^{-1} \, \tilde{\mb{M}}_F
	\end{array}
	\right ]
	\left [ \begin{array}{c}
	\tilde{\mb{x}}_{q,\ell} \\ \bm{\zeta}_{q,\ell}
	\end{array}
	\right ]
	\]
	which is stable if and only if condition (\ref{eq:rho}) holds.	
	\qed \vspace{.3cm}

       	Recall that, in view of the {\em Dahlquist's Equivalence Theorem}, zero-stability is necessary and sufficient for convergence of a 
	consistent time-discretization scheme \cite{num:stab}. Hence, under condition (\ref{eq:rho}), the hybrid Euler time-discretization scheme (\ref{time-discr:m}) turns out to be convergent.
	For instance, this means that in each interval $[t_q, t_{q+1}]$ the predicted estimates obtained via the Parallel Schwarz Consensus step (\ref{statespace_local:pred}) converge to the solution
	of a centralized prediction equation of the form
	\[
		\tilde{\mb{M}} \, \dot{ \hat{\mb{x}} } + \tilde{\mb{S}} \, \hat{\mb{x}} = \tilde{\mb{u}} 
	\]
	as the time-discretization step $\delta$ goes to $0$, or equivalently as the number $L$ of consensus steps goes to infinity. \vspace{.3cm}
	
	\begin{remark}
	Taking into account the particular structure of the FE mass matrix $\mb{M}$, which is reflected in the sparse structure of  $\tilde{\mb{M}}$, the numerical stability condition (\ref{eq:rho})
	is usually satisfied in practice (see, for instance, the simulation example of Section VI). In addition, in the unlikely case in which condition (\ref{eq:rho}) does not hold, 
	it is possible to modify the hybrid Euler time-discretization scheme (\ref{time-discr})
	(and hence the implementation of the Parallel Schwarz Consensus) so as to retrieve zero-stability. Specifically, by introducing a suitable  scalar $\omega \in ( 0, 1]$, one can replace (\ref{time-discr})
	with
	\begin{eqnarray}
	&&\tilde{\mb{M}}_D \left(\frac{\tilde{\mb{x}}_{q,\ell+1} - ( 2 - \omega )\, \tilde{\mb{x}}_{q,\ell} + (1-\omega) \,  \tilde{\mb{x}}_{q,\ell-1}  }{\omega \, \delta}\right) 
	\nonumber \\ && {} + \tilde{\mb{S}}_D\tilde{\mb{x}}_{q,\ell+1}  + \tilde{\mb{M}}_F \left(\frac{\tilde{\mb{x}}_{q,\ell}-\tilde{\mb{x}}_{q,\ell-1}}{\delta}\right) + \tilde{\mb{S}}_F\tilde{\mb{x}}_{q,\ell}  \nonumber \\ && {} =  
	\tilde{\mb{u}}_{q,\ell+1} + \tilde{\bm{\epsilon}}_{q,\ell+1} + \bm{\tau}_{q,\ell}
	\label{time-discr:omega}
	\end{eqnarray}
	which is still well-suited for distributed implementation.  Notice that such a modified scheme coincides with (\ref{time-discr}) for $\omega = 1$. Further,
	along the lines of Theorem 1, it is possible to show that (\ref{time-discr:omega}) is consistent for any value of $\omega \in (0,1]$, and zero-stable provided that
	\be
	\rho ( \omega \, \tilde{\mb{M}}_D^{-1} \, \tilde{\mb{M}}_F - (1 -\omega) \, \mb{I} )  <1 \, .
	\label{eq:rho:omega}
	\ee
	In turn, since 
	\[
	\rho ( \omega \, \tilde{\mb{M}}_D^{-1} \, \tilde{\mb{M}}_F - (1 -\omega) \, \mb{I} )  \le \max \{ \omega \,  \rho ( \tilde{\mb{M}}_D^{-1} \, \tilde{\mb{M}}_F ) , \,  1-\omega\}
	\]
	for any $\omega \in (0,1]$, condition (\ref{eq:rho:omega}) can be always satisfied
	for suitably small values of $\omega$ even 
	when condition (\ref{eq:rho}) does not hold. The price to be paid for the
	improved numerical stability is a slow-down of the information spread.
	\end{remark}

	\section{Stability analysis}

	In this section, the stability of the estimation error dynamics resulting from application of the distributed finite-element Kalman filter of Algorithm 2 is analyzed
	by supposing the measurement equation in each domain to be linear (as it happens when the sensors directly measure the target field like in (\ref{outm})). 
	Further, in order to simplify the notation, the interval $t_{q+1}-t_{q}$ between consecutive measurements is supposed to be constant, 
	so that in each sampling interval $[t_q , t_{q+1})$ a fixed number $L$ of consensus steps is performed.	
	With this respect,
	we make the following assumption. \vspace{.3cm}
	
	\begin{enumerate}[\bf {A}1.]
	\setcounter{enumi}{1}
	\item For each $m \in \mathcal N$, the local measurement function is linear, i.e., $\mb{h}^m (\mb{x}^m) = \mb{C}^m \mb{x}^m$. 
	Further, local observability holds in the sense that the pair $(({\mb A}^m)^L, \mb{C}^m)$ is observable for any $m \in \mathcal N$.
	\end{enumerate}
	\vspace{.3cm}
	
	Notice that the observability condition can be satisfied by choosing each subdomain large enough so that a sufficient number of sensors is included inside.
		
	Let us first rewrite (\ref{time-discr})
	into the state-space form
	\begin{eqnarray}
	\tilde{\mb{x}}_{q,\ell+1} &=& \underbrace{ \left( \tilde{\mb{M}}_D + \delta  \tilde{\mb{S}}_D \right)^{-1} \tilde{\mb{M}}_D }_{\tilde{\mb{A}}_D} \tilde{\mb{x}}_{q,\ell}    \nonumber \\ && {} 
	+ \underbrace{ \left( \tilde{\mb{M}}_D + \delta  \tilde{\mb{S}}_D \right)^{-1}
		\left( - \delta \tilde{\mb{S}}_F - \tilde{\mb{M}}_F \right) }_{\tilde{\mb{A}}_F} \tilde{\mb{x}}_{q,\ell}   \nonumber \\ && {} 
	+ \underbrace{ \left( \tilde{\mb{M}}_D + \delta  \tilde{\mb{S}}_D \right)^{-1} \tilde{\mb{M}}_F }_{\bar{\mb{A}}_F}  \tilde{\mb{x}}_{q,\ell-1} \nonumber \\ && {} 
	+ \underbrace{ \left( \tilde{\mb{M}}_D + \delta  \tilde{\mb{S}}_D \right)^{-1} \delta }_{\tilde{\mb{B}}} \, \tilde{\mb{u}}_{q,\ell+1} + \tilde{\mb{w}}_{q,\ell}
	\label{statespace_tilde}
	\end{eqnarray}
	where, clearly, $\tilde{\mb{A}}_D = {\rm block-diag} ( \mb{A}^{1},\dots,\mb{A}^{N} )$ is the block diagonal matrix of state transition matrices, representing the $N$ isolated subsystems.
	
	Recalling that, in each interval $[t_q, t_{q+1})$, the recursion (\ref{statespace_tilde}) is initialized with the initial conditions (\ref{eq:init:tilde}), it can be easily noticed that
	at the last consensus step $\ell = L$ one obtains
	\be
		\tilde{\mb{x}}_{q,L} = \tilde{\mb{A}}_D^{L} \, \tilde{\mb{x}}_{q,0} +  \tilde{\mb{A}}_{F,L} \tilde{\mb{x}}_{q,0} +  \tilde{\mb{B}}_{L} \tilde{\mb{U}}_{q} +
		 \tilde{\mb{D}}_{L} \tilde{\mb{W}}_{q} 
	\ee
	where $\tilde{\mb{U}}_{q} \defi col \{ \mb{u}_{q,\ell}, \, \ell = 1,\dots, L \}$,  $\tilde{\mb{W}}_{q} \defi col \{ \mb{w}_{q,\ell}, \, \ell = 1,\dots, L \}$
	and $\tilde{\mb{B}}_{L}$, $\tilde{\mb{D}}_{L}$, and $\tilde{\mb{A}}_{F,L}$ are suitable matrices 
	with the latter defining the interconnection couplings between subsystems. Noting that, by definition, $\tilde{\mb{x}}_{q,L} = \tilde{\mb{x}}_{q+1,0} = \tilde{\mb{x}} (T_{q+1} \Delta)$, the latter equation
	can be rewritten as 
	\be
		\tilde{\mb{x}}_{q+1} = \tilde{\mb{A}}_D^{L} \, \tilde{\mb{x}}_{q} +  \tilde{\mb{A}}_{F,L} \tilde{\mb{x}}_{q} +  \tilde{\mb{B}}_{L} \tilde{\mb{U}}_{q} +
		 \tilde{\mb{D}}_{L} \tilde{\mb{W}}_{q} 
	\ee
	where $\tilde{\mb{x}}_{q} \defi  \tilde{\mb{x}} (T_{q} \Delta)$.

	Similarly, application of step 3 of Algorithm 2 yields, at the last consensus step $\ell = L$,
	\be \label{eq:est1}
		\hat{\mb{x}}_{q,L} = \tilde{\mb{A}}_D^{L} \, \hat{\mb{x}}_{q,0} +  \tilde{\mb{A}}_{F,L} \hat{\mb{x}}_{q,0} +  \tilde{\mb{B}}_{L} \tilde{\mb{U}}_{q} \, .	
	\ee
	where $\hat{\mb{x}}_{q,\ell}  \defi col \{ \hat{\mb{x}}^m_{q,\ell} , \, m \in \mathcal N \}$. Further, by defining $\hat{\mb{x}}_{q|q}  \defi col \{ \hat{\mb{x}}^m_{q|q} , \, m \in \mathcal N \}$
	and $\hat{\mb{x}}_{q|q-1}  \defi col \{ \hat{\mb{x}}^m_{q|q-1} , \, m \in \mathcal N \}$, the global correction step of Algorithm 2 at time $t_{q+1}$ can be
	written as 
	\be \label{eq:est2}
	\hat{\mb{x}}_{q+1|q+1} = \hat{\mb{x}}_{q+1|q} + \tilde{\mb{L}}_{q+1} (  \tilde{\mb{y}}_{q+1} -  \tilde{\mb{C}}  \,   \hat{\mb{x}}_{q+1|q}     )
	\ee
	where $\tilde{{\mb{y}}}_{q+1}  \defi col \{ {\mb{y}}^m_{q+1} , \, m \in \mathcal N \}$, 
		   $\tilde{\mb{L}}_{q+1} = {\rm block-diag} ( \mb{L}^{1}_{q+1},\dots,\mb{L}^{N}_{q+1} )$, and
		   $ \tilde{{\mb{C}}}  \defi col \{ {\mb{C}}^m , \, m \in \mathcal N \} $.
		   
	Recalling that $\hat{\mb{x}}_{q,L} = \hat{\mb{x}}_{q+1|q}$ and $\hat{\mb{x}}_{q,0} = \hat{\mb{x}}_{q|q} $, equations (\ref{eq:est1}) and (\ref{eq:est2})  can be easily combined so as to write 
	$\hat{\mb{x}}_{q+1|q+1}$ as a function of $\hat{\mb{x}}_{q|q}$ so as to obtain a recursive expression for the global estimate. In addition, noting that the global output vector can be written as
	$\tilde{\mb{y}}_{q+1} = \tilde{\mb{C}}  \tilde{\mb{x}}_{q+1}  + \tilde{\mb{v}}_{q+1}$
	with $\tilde{{\mb{v}}}_{q+1}  \defi col \{ {\mb{v}}^m_{q+1} , \, m \in \mathcal N \}$, 
	we can also write a recursive expression for  the dynamics of the global estimation error $\tilde{\mb{e}}_{q} \defi col \{  \tilde{\mb{x}}_{q} - \hat{\mb{x}}_{q|q} , \, m \in \mathcal N \} $. Specifically,
	standard calculations yield
	\be \label{eq:est-err}
	\tilde{\mb{e}}_{q+1} = \left ( \mb{I} - \tilde{\mb{L}}_{q+1} \tilde{\mb{C}} \right ) \left ( \tilde{\mb{A}}_D^{L} + \tilde{\mb{A}}_{F,L}  \right )  \tilde{\mb{e}}_{q} + \tilde{\bm{\nu}}_q
	\ee
	where the term $\tilde{\bm{\nu}}_q =  ( \mb{I} - \tilde{\mb{L}}_{q+1} \tilde{\mb{C}}  )  \tilde{\mb{D}}_{L} \tilde{\mb{W}}_{q}  + \tilde{\mb{v}}_{q+1}$  accounts for the time/space discretization errors, for the
	measurement noise, and for all the other possible uncertainties.
	
	As for the time evolution of the global covariance matrix $ \tilde{\mb{P}}_{q|q} \defi {\rm block-diag} ( \mb{P}^{1}_{q|q},\dots,\mb{P}^{N}_{q|q} ) $, with similar reasoning as above
	it is an easy matter to see that application of Algorithm 2 leads to the following recursion
	\begin{eqnarray}
		\tilde{\mb{P}}_{q+1|q+1} &=& \left ( \mb{I}  - \tilde{\mb{L}}_{q+1} \tilde{\mb{C}}^T \right ) \tilde{\mb{P}}_{q+1|q} \nonumber \\
		&& \left ( \mb{I}  - \tilde{\mb{L}}_{q+1} \tilde{\mb{C}}^T \right ) \left [ \gamma^{2 L}  \tilde{\mb{A}}_D^{L} \tilde{\mb{P}}_{q|q} (\tilde{\mb{A}}_D^{L}  )^T +  \tilde{\mb{\Phi}} \right ] \nonumber \\
		\label{eq:cov}
	\end{eqnarray}
	where $ \tilde{\mb{\Phi}} \defi \sum_{i=0}^{L-1}  \gamma^{2 i}  \tilde{\mb{A}}_D^{i} \tilde{\mb{Q}} (\tilde{\mb{A}}_D^{i}  )^T $ and  $ \tilde{\mb{Q}} \defi {\rm block-diag} ( \mb{Q}^{1},\dots,\mb{Q}^{N} ) $.
	
	The following stability result can now be stated.
	 \vspace{.3 cm}
	
	\begin{theorem}
	Let assumptions A1 and A2 hold and let the matrices $\tilde{\mb{Q}}$ and $\tilde{\mb{R}} \defi {\rm block-diag} ( \mb{R}^{1},\dots,\mb{R}^{N} ) $ be positive definite. 
	Then, the global covariance matrix asymptotically converges to the the unique positive solution $\tilde{\mb{P}}$ of the
	algebraic Riccati equation
	\[
	\tilde{\mb{P}}^{-1} = \left [ \gamma^{2 L}  \tilde{\mb{A}}_D^{L} \tilde{\mb{P}} (\tilde{\mb{A}}_D^{L}  )^T +  \tilde{\mb{\Phi}} \right ]^{-1} + \tilde{\mb{C}}^T \, \tilde{\mb{R}}^{-1} \, \tilde{\mb{C}} \, .
	\]
	
	In addition, if the scalar $\gamma$ is chosen so that
	\be \label{eq:stab}
	\gamma^L > \left \| \mb{I} + \left ( \tilde{\mb{A}}_D^{L} \right )^{-1}  \tilde{\mb{A}}_{F,L}  \right \|_{\tilde{\mb{P}}} \, ,
	\ee
	where $\| \cdot \|_{\mb{M}}$ denotes the matrix norm induced by the vector norm 
	$\| \mb{x} \|_{\mb{M}} \defi \sqrt{\mb{x}^T
	\mb{M} \mb{x}}$,
	then the dynamics (\ref{eq:est-err}) of the estimation error is exponentially stable.
	\end{theorem} \vspace{.3cm}
	{\em Proof:} Notice first that assumption A2 implies observability of the pair $(\tilde{\mb{A}}_D^L, \tilde{\mb{C}} )$ which, as it can be easily verified through the PBH test, also implies observability of 
	$(\gamma^L \tilde{\mb{A}}_D^L, \tilde{\mb{C}} )$ for any real $\gamma >0$.	
	Then, the convergence of  $\tilde{\mb{P}}_{q|q}$ to $\tilde{\mb{P}} >0$ follows from well known results on discrete-time Kalman filtering, since (\ref{eq:cov}) is
	the standard Kalman filter covariance recursion for a linear system with state matrix $\gamma^L \tilde{\mb{A}}_D^L$ and output matrix $\tilde{\mb{C}}$.
	
	Let now $\tilde{\mb{L}}$ be the steady-state global Kalman gain associated with the steady-state covariance $\tilde{\mb{P}}$. With standard manipulations, it can be seen that $\tilde{\mb{L}}$ and $\tilde{\mb{P}}$
	satisfy the relationship
	\[
	\tilde{\mb{P}} =  ( \mb{I} - \tilde{\mb{L}} \tilde{\mb{C}}^T  ) \left [ \gamma^{2 L}  \tilde{\mb{A}}_D^{L} \tilde{\mb{P}} (\tilde{\mb{A}}_D^{L}  )^T +  \tilde{\mb{\Phi}} \right ] 
	( \mb{I} - \tilde{\mb{L}} \tilde{\mb{C}}^T  )^T + \tilde{\mb{L}}  \tilde{\mb{R}} \tilde{\mb{L}}^T
	\]
	so that
	\[
	( \mb{I} - \tilde{\mb{L}} \tilde{\mb{C}}^T  ) \left [ \gamma^{2 L}  \tilde{\mb{A}}_D^{L} \tilde{\mb{P}} (\tilde{\mb{A}}_D^{L}  )^T  \right ]  ( \mb{I} - \tilde{\mb{L}} \tilde{\mb{C}}^T )^T \le   \tilde{\mb{P}} 
	\]
	and, hence, 
	\be \label{eq:proof:stab}
	\left \| ( \mb{I} - \tilde{\mb{L}} \tilde{\mb{C}}^T  )  \tilde{\mb{A}}_D^{L}  \right \|_{\tilde{\mb{P}}}  \le 1/ \gamma^L \, .
	\ee
	Notice now that the matrix $ \left ( \mb{I} - \tilde{\mb{L}}_{q+1} \tilde{\mb{C}} \right ) \left ( \tilde{\mb{A}}_D^{L} + \tilde{\mb{A}}_{F,L}  \right ) $, which determines the dynamics of the estimation error,
	exponentially converges to $ \left ( \mb{I} - \tilde{\mb{L}} \tilde{\mb{C}} \right ) \left ( \tilde{\mb{A}}_D^{L} + \tilde{\mb{A}}_{F,L}  \right ) $, so that the estimation error dynamics is exponentially stable
	if and only if $ \left ( \mb{I} - \tilde{\mb{L}} \tilde{\mb{C}} \right ) \left ( \tilde{\mb{A}}_D^{L} + \tilde{\mb{A}}_{F,L}  \right ) $ is Schur stable.
	Hence, in order to complete the proof, it is sufficient to observe that
	\begin{eqnarray*}
	\lefteqn{ \left \|  \left ( \mb{I} - \tilde{\mb{L}} \tilde{\mb{C}} \right ) \left ( \tilde{\mb{A}}_D^{L} + \tilde{\mb{A}}_{F,L}  \right ) \right \|_{\tilde{\mb{P}}} } \\ 
	&& {} \le
	\bigg \|  \left ( \mb{I} - \tilde{\mb{L}} \tilde{\mb{C}} \right ) \, \tilde{\mb{A}}_D^{L}  \bigg  \|_{\tilde{\mb{P}}} \,
	\bigg \|  \mb{I} + \left ( \tilde{\mb{A}}_D^{L} \right )^{-1}  \tilde{\mb{A}}_{F,L}  \bigg \|_{\tilde{\mb{P}}} \\
	&&  {} \le  \bigg \|  \mb{I} + \left ( \tilde{\mb{A}}_D^{L} \right )^{-1}  \tilde{\mb{A}}_{F,L}  \bigg \|_{\tilde{\mb{P}}} \,  /\gamma^L
	\end{eqnarray*} 
	where the latter inequality follows from (\ref{eq:proof:stab}). In fact, this implies that 
	$\left \|  \left ( \mb{I} - \tilde{\mb{L}} \tilde{\mb{C}} \right ) \left ( \tilde{\mb{A}}_D^{L} + \tilde{\mb{A}}_{F,L}  \right ) \right \|_{\tilde{\mb{P}}}  < 1$ whenever (\ref{eq:stab}) holds.
	\qed \vspace{.3cm}

	\section{Simulation Experiments}
	
	This section provides numerical examples and relative results illustrating the effectiveness of the proposed distributed \textit{finite element Kalman filter} presented in section IV. Consider the transient \textit{heat conduction} problem, introduced in section II as a particular example of \eqref{PDE}, in a thin polygonal metal plate with constant, homogeneous, and isotropic properties. Assuming the thickness of the slab is considerably smaller than the planar dimensions, then the temperature can be assumed to be constant along the width direction, and the problem is reduced to two dimensions.
	Hence, the diffusion process in a thin plate is modelled by the 2D parabolic PDE $\, {\partial x} / {\partial t} - \lambda \left ( {\partial^2 x} / {\partial \xi^2} + {\partial^2 x} / {\partial \eta^2} \right ) = 0$ with boundary condition $\mathcal B (x) = \alpha(\xi,\eta) \,  {\partial x}/{\partial \mb{n}} + \beta(\xi,\eta) x$ such that $\alpha(\xi,\eta) \,\beta(\xi,\eta) \ge 0 $, $\alpha(\xi,\eta) + \beta(\xi,\eta) > 0, \, \forall (\xi,\eta) \in \partial \Omega$. Notice that, $x(\xi,\eta,t)$ denotes the temperature as a function of time $t$ and spatial variables $(\xi,\eta) \in \Omega$, $f = 0$ stands for no inner heat-generation, whereas $\lambda = 1.11 \times 10^{-4} \left[m^2/s\right]$ is the \textit{thermal diffusivity} of copper at $25 \, [\degree C]$ (Table 12, Appendix 2 in \cite{Kreith10}), assumed to be constant in time and space.
	
\begin{figure}[!t]
					\centering
					\includegraphics[width=.5\columnwidth]{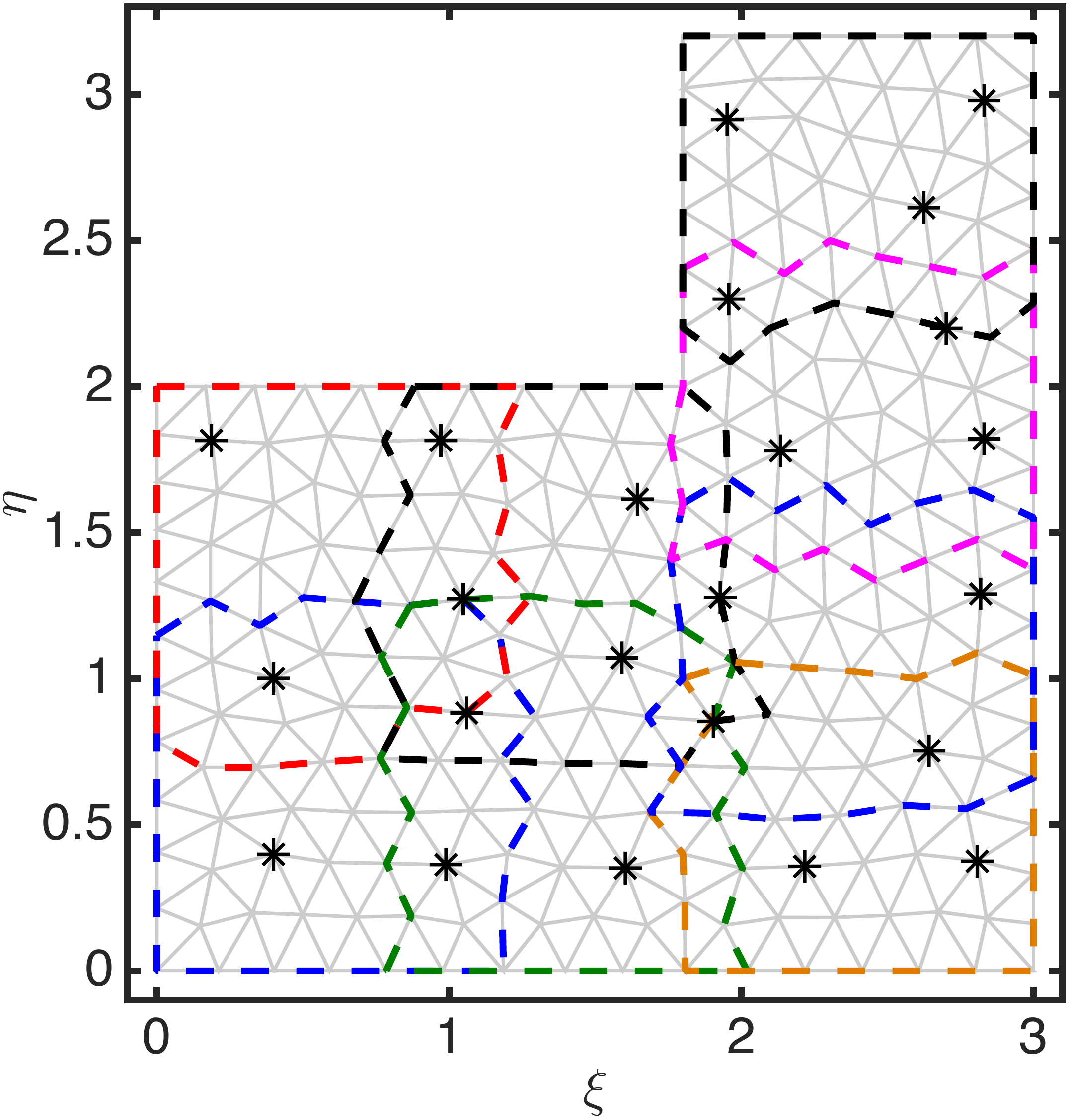}
					\caption{Global FE mesh (grid of solid lines) generated over $\Omega$ and domain decomposition into 8 overlapping subdomains (dashed polygons). The position of each sensor is denoted by $\ast$.}
					\label{fig:subdomains}
	\end{figure}

	A network of $S = 23$ sensors (Fig. \ref{fig:subdomains}) located in the known positions $\mb{s}_i = \left[ \xi_i,\eta_i \right]^T$ is assumed to collect point temperature measurements at regularly time-spaced instants $t_q = q\,T_s$, with $T_s = 100 \, [s]$ and standard deviation of measurement noise $\sigma_v = 0.1 \, [K]$. The considered sensor network has been chosen to guarantee local observability (assumption A2).
	
	The MATLAB PDE Toolbox is used to generate the triangular mesh (252 vertices, 436 elements) shown in Fig. \ref{fig:subdomains} of size $b = 0.2 [m]$ (defined as the length of the longest edge of the element), over the global 2D domain $\Omega$.
	Next, as can be seen from Fig. \ref{fig:subdomains}, the domain under consideration is decomposed into $N = 8$ overlapping subdomains $\Omega_m$, i.e. 
	$\mathcal{N} = \{ 1, \dots, 8 \}$, each being assigned to a \textit{node} with local processing and communication capabilities. 
	It is worth pointing out that domain decomposition comes with an appropriate partitioning of the original global mesh so that the resulting local grids actually match on the regions of overlap between subdomains.
	
	Domain triangulation allows for a simple construction of basis functions $\{\phi_{j} (\xi,\eta)\}_{j=1}^n$, 
	which are continuous piecewise polynomial functions, such that their value is unity in vertex $j$ and vanishes at the remaining vertices, i.e.
	\begin{eqnarray*}
		&& \hspace{-.8cm} \phi_j(\xi_i,\eta_i) = \left \{
		\ba{ll}
		1 & \mbox{ if } i = j ~~~~~~~~~~~~~~~i,j = 1,2,...,n\\
		0   & \mbox{ if } i \neq j \\
		\ea \right . \nonumber \\
	\end{eqnarray*}
	Here we use continuous piecewise linear functions defined on each element as $\psi_{\mathcal{E}}(\xi,\eta) = c_0 + c_1\xi + c_2\eta$ with $(\xi,\eta) \in \mathcal{E}$ and $c_0, c_1, c_2 \in \rr$, so that each function is uniquely determined by its three nodal values $x_i = \psi_{\mathcal{E}}(\xi_i,\eta_i)$, $i \in {\mathcal{E}}$.
	
	Basis functions are used off-line by the FE centralized filter and in the distributed setup for the element-by-element construction of matrices $\mb{S}$ and $\mb{M}$, introduced in \eqref{WF}.
	Then, the state dynamics of the centralized filter can be directly computed, whereas local estimators first need to extract matrices $\mb{M}^{mm}, \mb{S}^{mm}$ and $\mb{M}^{mj}$, $\mb{S}^{mj}$ in order to calculate $\mb{A}^m, \mb{A}^{m j}$ and $\bar{\mb{A}}^{m j}$ which finally provide the finite-dimensional model of temperature evolution in $\Omega_m$ through \eqref{statespace_local}. 
	Notice that these matrices are evaluated for a fixed sampling interval $\delta = T_s / L$, where $L$ denotes the number of consensus iterations $L_q$ introduced in Section IV, here assumed constant in each sampling interval $q$.
	For a fair comparison between centralized and distributed approaches, a constant time integration interval $\Delta = 10 \, [s]$ has been chosen for the centralised filter.
	
\begin{figure}[!t] 
    			\centering 
    			\subfloat[$\mb{S}$: 1626 nonzero elements]{\includegraphics[width=.5\columnwidth]{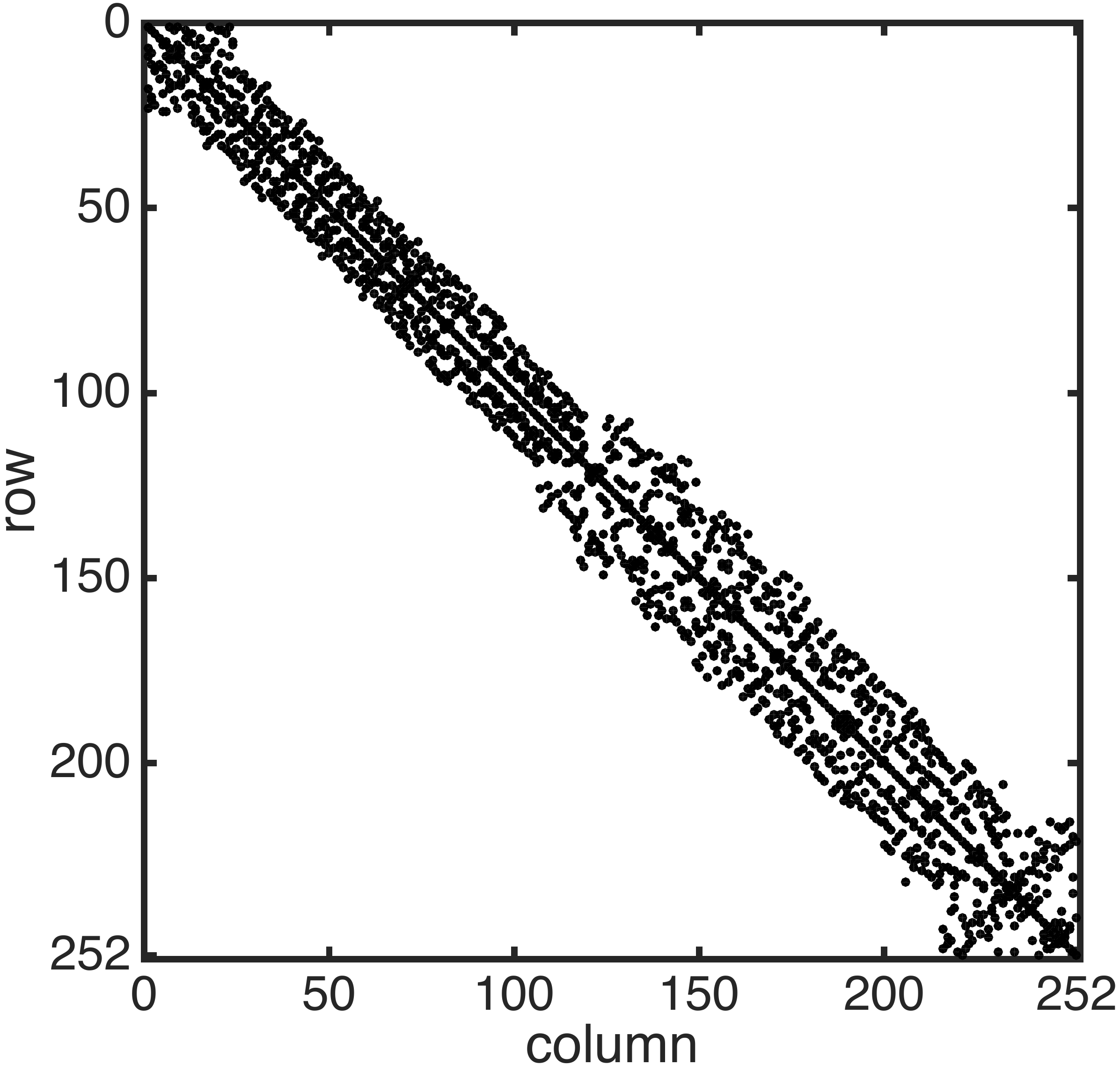}%
    				\label{fig:S_centr}}
    			\hfil
    			\subfloat[$\tilde{\mb{S}}_D$: 1632 nonzero elements (red); $\tilde{\mb{S}}_F$: 223 nonzero elements (blue)]{\includegraphics[width=.5\columnwidth]{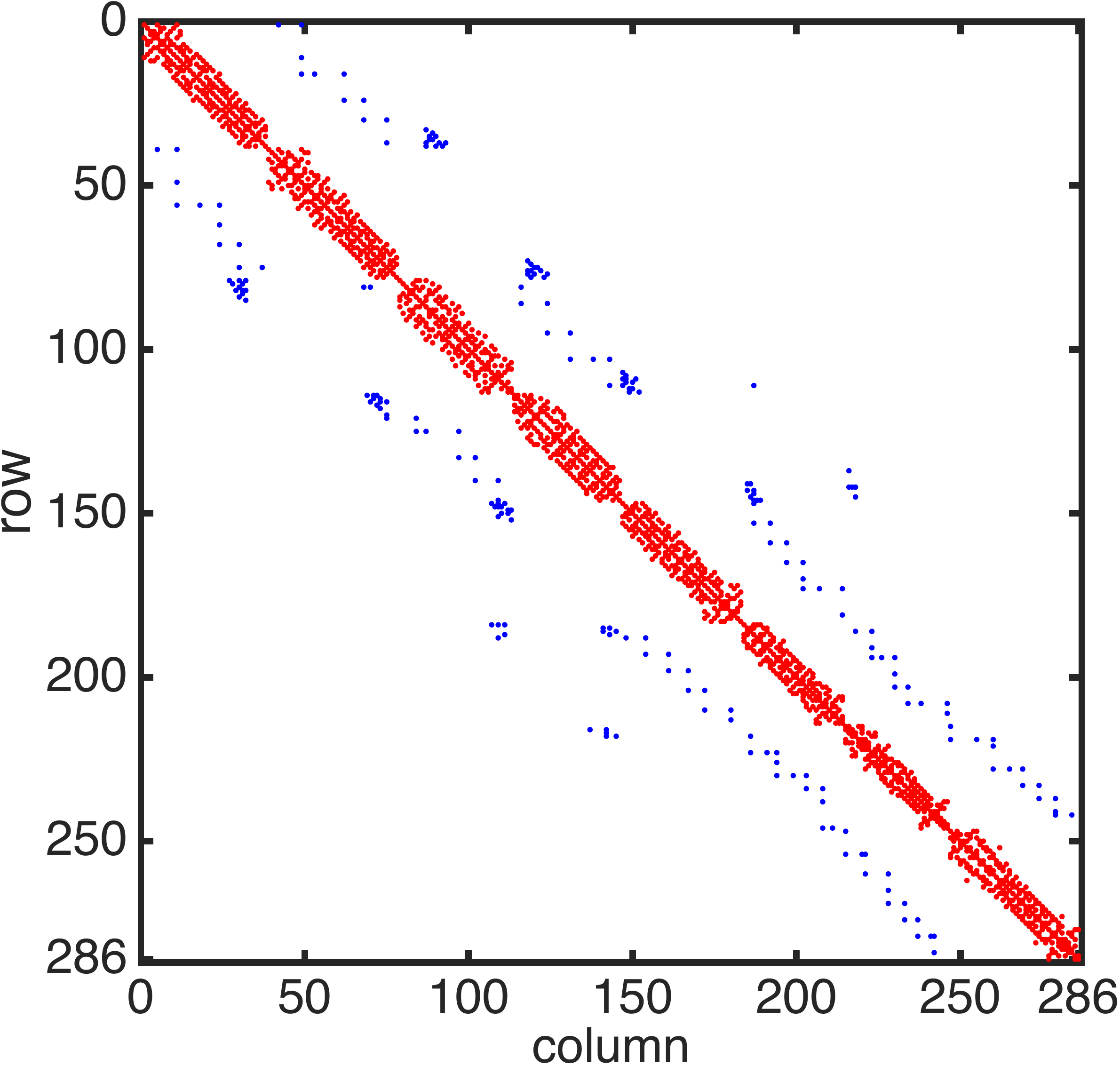}%
    				\label{fig:SD_SF}}
    			\caption{Sparsity pattern of $252 \times 252$ matrix $\mb{S}$ (a), and $286 \times 286$ matrix $\tilde{\mb{S}} = \tilde{\mb{S}}_D + \tilde{\mb{S}}_F$ (b).}
    			\label{fig_spyS}
    		\end{figure}

\begin{figure}[!t]
    				\centering
    				\includegraphics[width=.5\columnwidth]{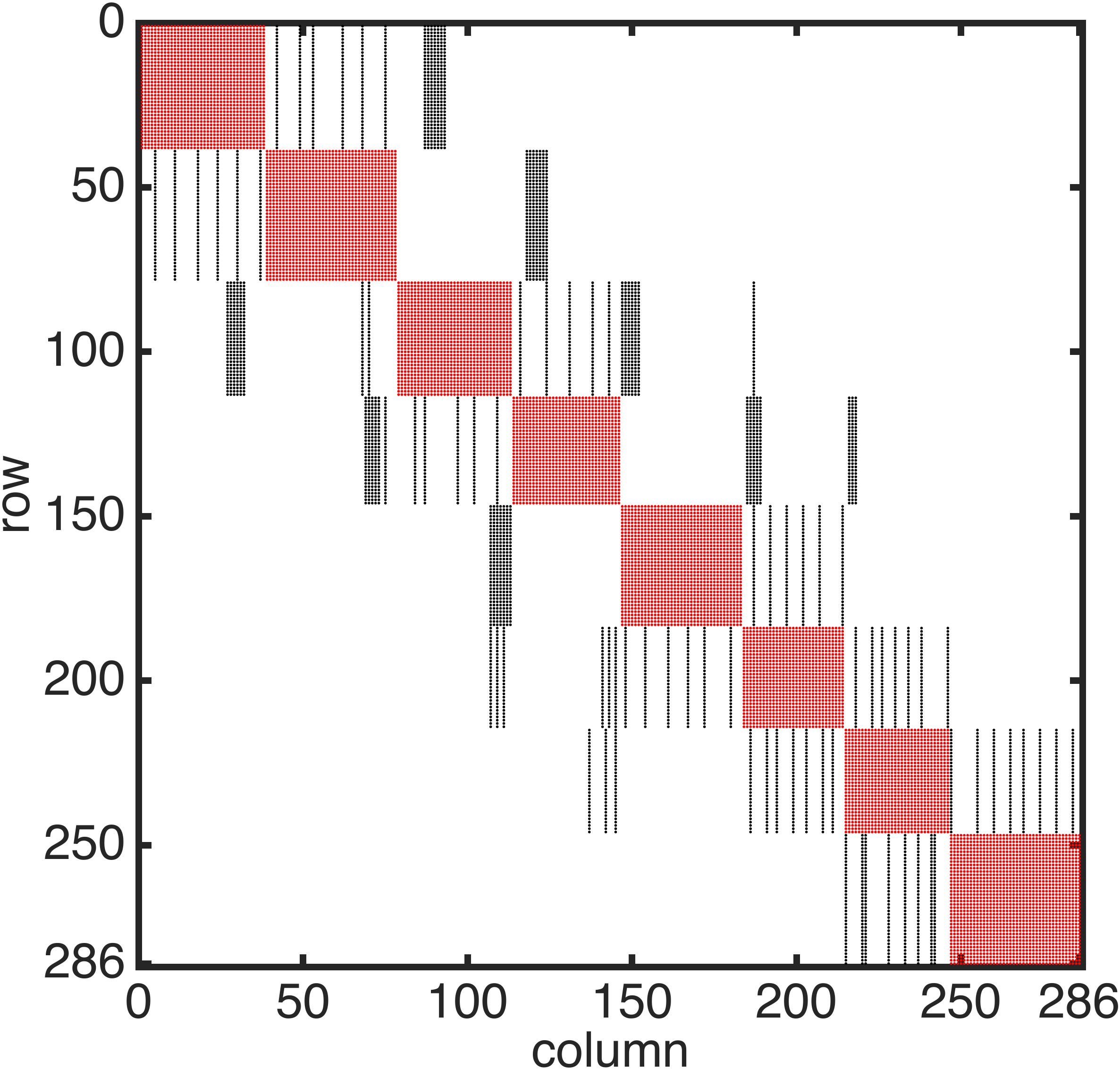}
    				\caption{Sparsity pattern of $\tilde{\mb{A}}_D$ (red) and $\tilde{\mb{A}}_F$ (black).}
    				\label{fig:AD_AF}
    			\end{figure}

	Notice that, being $\{\phi_{j} (\xi,\eta)\}_{j=1}^n$ functions with a small support defined by the set of triangles sharing node $j$, the resulting \textit{mass} and \textit{stiffness} matrices will be sparse, with the same pattern shown in Fig. \ref{fig:S_centr}.
	In Fig. \ref{fig:SD_SF} it can be seen how the structure of the stiffness matrix changes when considering the augmented system \eqref{eq:diffeqtilde}.
	The distributed pattern of the networked system is highlighted in Fig. \ref{fig:AD_AF}, where $\tilde{\mb{A}}_D$ represents each subsystem as isolated, though affected by the evolution of neighbors through $\tilde{\mb{A}}_F$.
    		
	In the following experiments, both FE filters assume the initial temperature field of the plate uniform at $x_0(\xi,\eta) = 300 \, [K]$, and the a-priori estimate taken as first guess $\hat{x}_{1|0}(\xi,\eta) = 305 \, [K]$, with diagonal covariance $\mb{P}_{1|0} = 20 \, \mb{I}$. Moreover, a zero-mean white noise process has been assumed, with covariance $\mb{Q} = \sigma_w^2 \, \mb{I}$, where $\sigma_{w} = 3 \, [K]$.
	Taking into consideration model uncertainty, the \textit{ground truth} of the experiments is represented by a real process simulator implementing a finer mesh (915 vertices, 1695 elements) of size $b = 0.1$ instead of $b=0.2$, running at a higher sample rate ($1 \, Hz$), and aware of the possibly time-varying boundary conditions of the system.
	On the other hand, both distributed and centralised filters have no knowledge of the real system boundary conditions, so they simply assume the plate adiabatic on each side.
	
	The performance of the novel distributed \textit{FE Kalman filter} has been evaluated in terms of Root Mean Square Error (RMSE) of the estimated temperature field, averaged over a set of about 300 sampling points uniformly spread within the domain $\Omega$, and 500 independent Monte Carlo realizations.
		\begin{figure}[!t]
			\centering
			\includegraphics[width=\columnwidth]{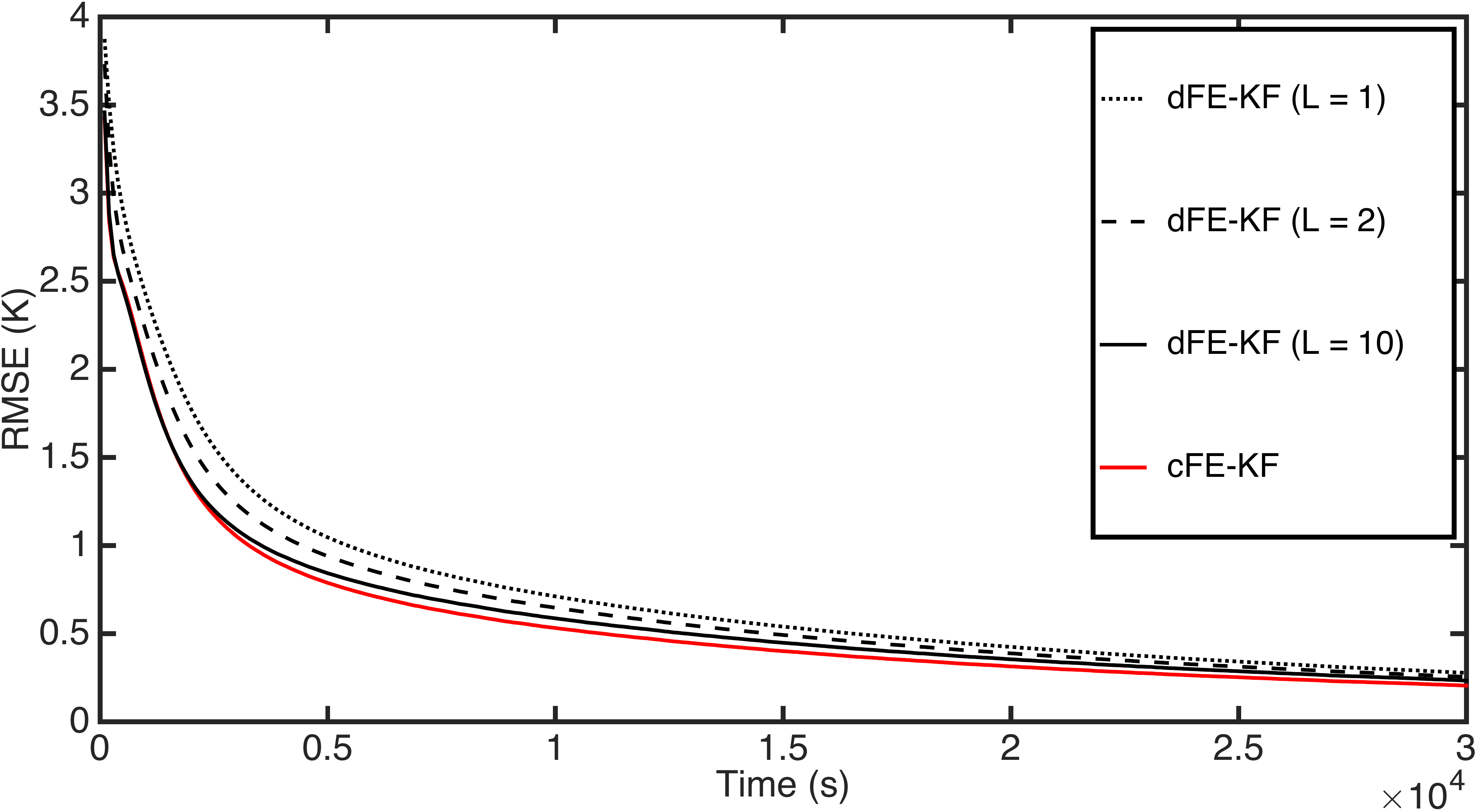}
			\caption{Scenario 1: Comparison of performance of centralised and distributed FE-KF ($\gamma = 1.1$).}
			\label{fig:rmse1}
		\end{figure}
		\begin{figure}[t!]
			\centering
			\includegraphics[width=\columnwidth]{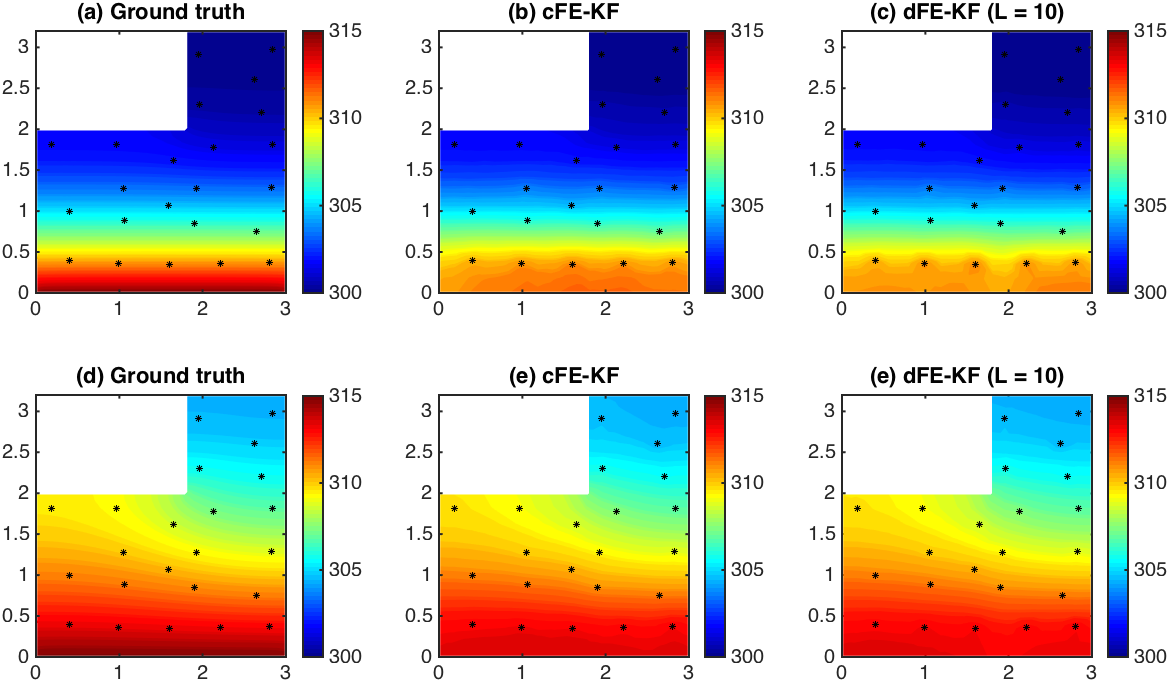}
			\caption{Scenario 1: \textit{True} and estimated temperature fields at time steps $q = 50$ (a,b,c) and $q = 200$ (d,e,f).}
			\label{fig:field1}
		\end{figure}  
	\subsection*{Scenario 1}
\begin{figure}[!t]
					\centering
					\includegraphics[width=\columnwidth]{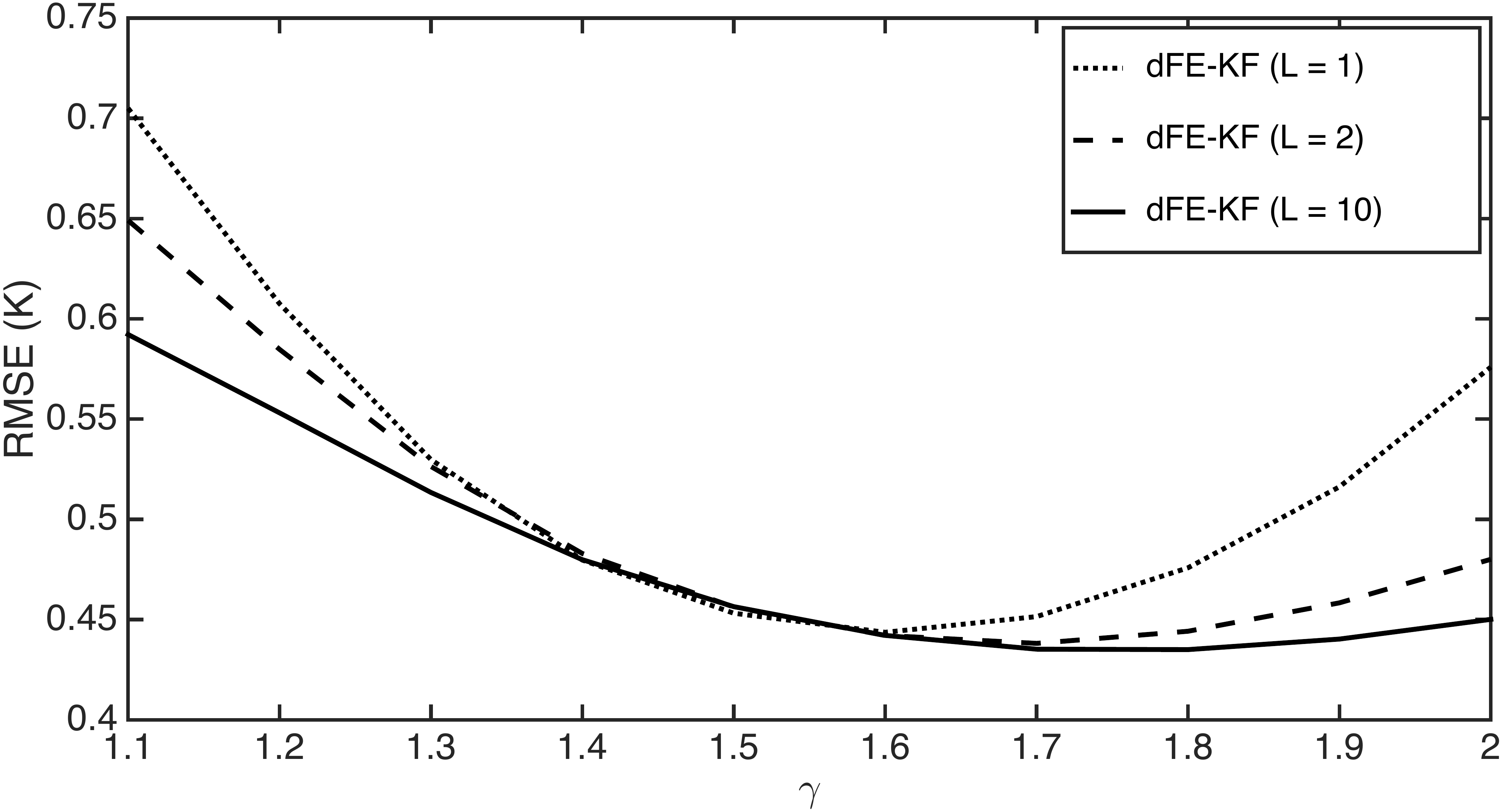}
					\caption{Scenario 1: Comparison of the mean value of the RMSE for different values of $\gamma$.}
					\label{fig:rmse_gamma}
				\end{figure}

	In the first example, transient analysis is performed on a thin adiabatic L-shaped plate (seen in Fig. \ref{fig:subdomains}) with a fixed temperature along the bottom edge. This is a problem with mixed boundary conditions, namely a non-homogeneous Dirichlet condition on the bottom edge of the plate $\partial \Omega_1$, i.e. 
	\be
	x = T_1 \,\,\,\,\,\,\, \mbox{ on } \partial \Omega_1,
	\label{Dbc}
	\ee
	where $T_1 = 315 \, [K]$, and natural homogeneous Neumann boundary conditions on the remaining insulated sides $\partial \Omega_2 = \partial \Omega \setminus \partial \Omega_1$, so that
	\be
	{\partial x}/{\partial \mb{n}} = 0 \,\,\,\,\,\,\, \mbox{ on } \partial \Omega_2.
	\label{Nbc}
	\ee
					\begin{figure}[!t]
						\centering
						\includegraphics[width=\columnwidth]{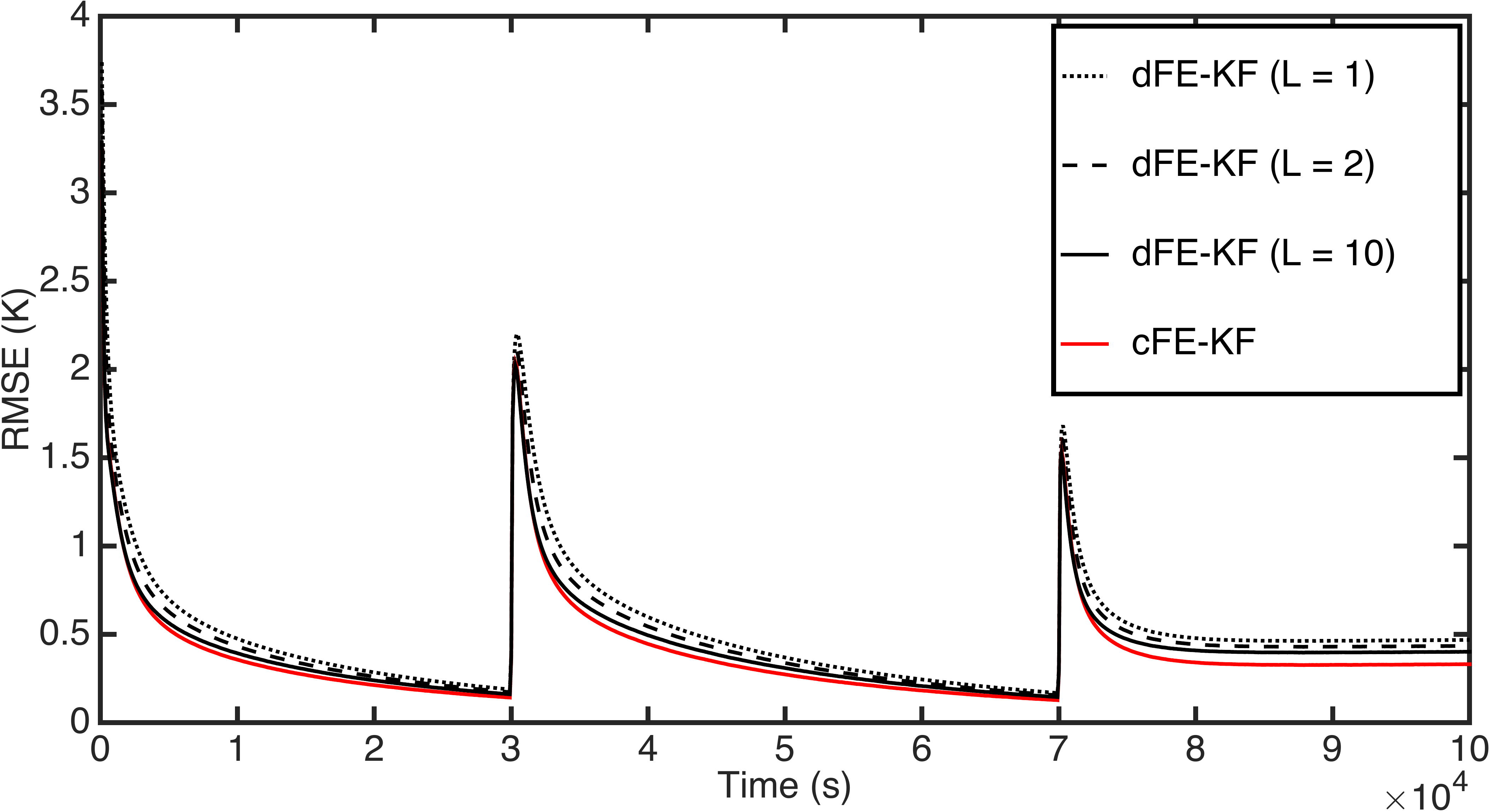}
						\caption{Scenario 2: Comparison of performance of centralised and distributed FE-KF ($\gamma = 1.1$).}
						\label{fig:rmse2}
					\end{figure}
					\begin{figure}[t!]
						\centering
						\includegraphics[width=\columnwidth]{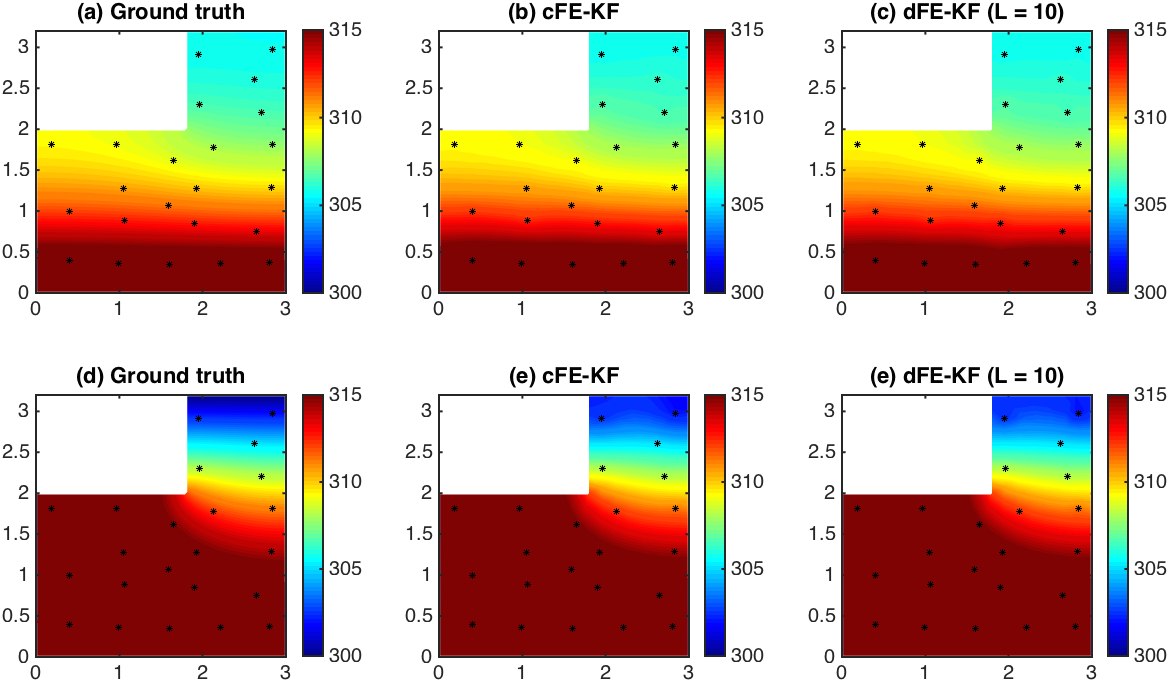}
						\caption{Scenario 2: \textit{True} and estimated temperature fields at time steps $q = 350$ (a,b,c) and $q = 900$ (d,e,f).}
						\label{fig:field2}
					\end{figure} 
	The duration of each Monte Carlo run is fixed to $3 \times 10^4 \, [s]$ (300 samples).  
	Fig. \ref{fig:rmse1} illustrates the performance comparison between centralized (cFE-KF) and distributed (dFE-KF) filters for $\gamma = 1.1$ and for three different values of the parameter $L$ adopted in the distributed framework. 
	First of all, it can be seen that both FE algorithms succeed in reconstructing the \textit{true} field of the system based on fixed, point-wise temperature observations.
	Moreover, the performance of the distributed FE filters is very close, even for $L = 1$, to that of the centralized filter, which collects all the data in a central node. 
	Last but not least, in the distributed setting the RMSE behaviour improves by increasing the number of consensus steps. 
	This is true for certain values of $\gamma$, whereas for others the difference in performance is considerably reduced, as clearly presented in Fig. \ref{fig:rmse_gamma}.  
	Note that the covariance boosting factor used in (\ref{LDTM-cov}) is set to 
	$\gamma_L = \sqrt[L]{\gamma}, \, \forall L = 1,2,10$, in order to obtain a fairly comparable effect of covariance inflation after $L$ consensus steps for different distributed filters.   
	Further insight on the performance of the proposed FE estimators is provided in Fig. \ref{fig:field1}, which shows the \textit{true} and estimated temperature fields at two different time steps $q = 50$ and $q = 200$, obtained in a single Monte Carlo experiment by using cFE-KF and dFE-KF with $L=10$. 
	
	\subsection*{Scenario 2}
	
	In the second experiment, two time-varying disturbances have been added in order to test the robustness of the proposed FE estimators in a more challenging scenario. To this end, different boundary conditions are considered. Specifically, a time-dependent Dirichlet condition \eqref{Dbc} with $T_1 = 310 \, [K]$ for time steps $q \in \{0,...,299\}$, and $T_2 = 320 \, [K]$ for $q \in \{300,...,1000\}$, is set on all nodes of the bottom edge $\partial \Omega_1$. The top edge of the plate $\partial \Omega_3$ is first assumed adiabatic for $q \in \{0,...,699\}$, then the inhomogeneous Robin boundary condition 
	\be
	\lambda \, {\partial x}/{\partial \mb{n}} + \nu \, x = \nu \, x_e \,\,\,\, \mbox{ on } \partial \Omega_3
	\label{Rbc}
	\ee
	is applied for $q \in \{700,...,1000\}$. This models a sudden exposure of the surface to a fluid, fixed at an external temperature $x_e = 300 \, [K]$, through a
	uniform and constant convection heat transfer coefficient $\nu = 10 \,[W/m^2 K]$. 
	The remaining edges $\partial \Omega_2$ where \eqref{Nbc} holds, are assumed thermally insulated for the duration of the whole experiment, lasting $10^5 \, [s]$ ($1000$ samples).
	
%
%
%
	
	Performance of the proposed distributed filter has been evaluated for different values of $L$ over 500 independent Monte Carlo runs and compared to the behavior of the centralized FE Kalman filter. Simulation results, in Fig. \ref{fig:rmse2}, show that the proposed FE estimators provide comparable performance to the centralized filter, moreover the gap reduces as $L$ increases. It is worth pointing out that the peaks appearing in the RMSE plot, displayed in Fig. \ref{fig:rmse2}, are due to the abrupt changes of the unknown boundary conditions, which cause considerable jumps of the estimation errors at time steps $300$ and $700$.
	Nevertheless, the filters under consideration manage to compensate for the lack of knowledge and effectively reduce the error, even if, due to persistent and cumulative disturbances on the inferred field profile, errors do not converge to zero.
	The original \textit{ground truth} and the reconstructed fields are depicted in Fig. \ref{fig:field2} for $q = 350$ and $q = 900$.
	

	\section{Conclusions}
	The paper has dealt with the decentralised estimation of a time-evolving and space-dependent field governed by a linear partial differential equation, given point-in space measurements of multiple sensors deployed over the area of interest.
	The originally infinite-dimensional filtering problem has been approximated into a finite-dimensional large-scale one via the \textit{finite element} method and, further, a consensus approach inspired by the parallel Schwarz method for domain decomposition has allowed to nicely scale the
	overall problem complexity with respect to the number of used processing nodes.
	Combining these two ingredients, a novel computationally efficient consensus finite-element Kalman filter has been proposed to solve in a decentralized and scalable fashion filtering problems involving distributed-parameter systems.
Both numerical stability of the finite-element approximation and exponential stability of the proposed consensus finite-element Kalman filter have been analysed.
Simulation experiments have been presented in order to demonstrate the validity of the proposed approach.

The results of this work can be extended to the estimation of fields governed by more general partial differential equations and also
be applied to the estimation/localization of diffusive sources.


	
	%

	

	\ifCLASSOPTIONcaptionsoff
	\newpage
	\fi

\end{document}